\documentclass[%
 reprint,
superscriptaddress,
 amsmath,amssymb,
 aps,
prl,
]{revtex4-2}

\usepackage{comment}
\usepackage{graphicx}
\usepackage{dcolumn}
\usepackage{bm}
\usepackage[version=4]{mhchem}
\usepackage{chemformula}
\usepackage{siunitx}
\usepackage{hyperref}
\usepackage{booktabs}
\usepackage{makecell}
\usepackage{siunitx}
\usepackage{threeparttable}
\begin{document}

\preprint{dd in RIXS}

\title{Machine-learning test of the single-ion model for $dd$ excitations in cuprates}

\author{Maryia\, Zinouyeva}
\email[e-mail ]{maryia.zinouyeva@polimi.it}
\affiliation{Dipartimento di Fisica, Politecnico di Milano, piazza Leonardo da Vinci 32, I-20133 Milano, Italy}

\author{Leonardo\, Martinelli}
\affiliation{Dipartimento di Fisica, Politecnico di Milano, piazza Leonardo da Vinci 32, I-20133 Milano, Italy}
\affiliation{Physik-Institut, Universität Zürich, Winterthurerstrasse 190, CH-8057 Zürich, Switzerland}

\author{Riccardo\, Arpaia}
\affiliation{Department of Molecular Sciences and Nanosystems, Ca’ Foscari University of Venice, I-30172 Venezia, Italy}
\affiliation{
Quantum Device Physics Laboratory, Department of Microtechnology and Nanoscience, Chalmers University of Technology, SE-41296 Göteborg, Sweden
}


\author{Nicholas B.\, Brookes}
\affiliation{ESRF, The European Synchrotron, 71 Avenue des Martyrs, CS 40220, F-38043 Grenoble, France}

\author{Daniele\, Di Castro}
\affiliation{Dipartimento di Ingegneria Civile e Ingegneria Informatica, Università di Roma Tor Vergata, I-00133 Roma, Italy}
\affiliation{CNR-SPIN, Università di Roma Tor Vergata, I-00133 Roma, Italy}

\author{Kurt\, Kummer}
\affiliation{ESRF, The European Synchrotron, 71 Avenue des Martyrs, CS 40220, F-38043 Grenoble, France}

\author{Floriana\, Lombardi}
\affiliation{
Quantum Device Physics Laboratory, Department of Microtechnology and Nanoscience, Chalmers University of Technology, SE-41296 Göteborg, Sweden
}

\author{Giacomo\, Merzoni}
 \affiliation{Dipartimento di Fisica, Politecnico di Milano, piazza Leonardo da Vinci 32, I-20133 Milano, Italy}

 \author{Francesco\, Rosa}
 \altaffiliation[current address ]{Institute of Physics II, University of Cologne, 50937 Cologne, Germany}
\affiliation{Dipartimento di Fisica, Politecnico di Milano, piazza Leonardo da Vinci 32, I-20133 Milano, Italy}

\author{Alessandro\, Tarasio}
\affiliation{Dipartimento di Fisica, Università della Calabria, and Istituto Nazionale di Fisica Nucleare, Gruppo Collegato di Cosenza, Ponte Pietro Bucci Cubo 31C, I-87036 Rende, Italy}

\author{Enrico\, Tassi}
\affiliation{Dipartimento di Fisica, Università della Calabria, and Istituto Nazionale di Fisica Nucleare, Gruppo Collegato di Cosenza, Ponte Pietro Bucci Cubo 31C, I-87036 Rende, Italy}

\author{Flora\, Yakhou-Harris}
\affiliation{ESRF, The European Synchrotron, 71 Avenue des Martyrs, CS 40220, F-38043 Grenoble, France}

\author{Ezio\, Puppin}
\affiliation{Dipartimento di Fisica, Politecnico di Milano, piazza Leonardo da Vinci 32, I-20133 Milano, Italy}

\author{Marco\, Moretti Sala}
\affiliation{Dipartimento di Fisica, Politecnico di Milano, piazza Leonardo da Vinci 32, I-20133 Milano, Italy}

\author{Giacomo\, Ghiringhelli}
\email[e-mail ]{giacomo.ghiringhelli@polimi.it}
\affiliation{Dipartimento di Fisica, Politecnico di Milano, piazza Leonardo da Vinci 32, I-20133 Milano, Italy}
\affiliation{CNR-SPIN, Dipartimento di Fisica, Politecnico di Milano, I-20133 Milano, Italy}

\date{\today}

\begin{abstract}

We investigate $dd$ excitations in Resonant Inelastic X-ray Scattering spectra of YBa$_2$Cu$_3$O$_6$ and La$_2$CuO$_4$ using the local single-ion model. The data are analyzed by conventional global fitting and by a convolutional neural network trained within the same theoretical framework. For YBa$_2$Cu$_3$O$_6$, the excited state energies obtained with the two methods coincide, leading to the $xy$, $3z^2-r^2$, $xz/yz$ sequence for increasing energy. This result validates the use of machine learning tools for the analysis of RIXS spectra dominated by $dd$ excitations. By contrast, for La$_2$CuO$_4$, the two methods do not converge to a single solution, revealing the limitations of the single-ion model in describing $dd$ excitations in cuprates and pointing to the role of additional contributions beyond a purely local picture in shaping high-energy excitations.


\end{abstract}

\maketitle


Orbital degrees of freedom play an important role in transition-metal compounds, where they drive structural transitions via the cooperative Jahn-Teller effect and orbital order \cite{Murakami_PhysRevLett.81.582}, and determine the magnetic ground state \cite{Goodenough1955, Kanamori1959, Anderson1959} with its associated properties. In cuprates, $dd$ excitations between Cu $3d$ orbitals are of particular interest, since the energy splitting between the $3z^2-r^2$ and $x^2-y^2$ states is believed to be a key parameter controlling the superconducting transition temperature \cite{Pavarini_PhysRevLett.87.047003}. Moreover, several theoretical studies have pointed to a close connection between the Cu $3d$ electronic structure, the superexchange interaction \cite{bogdanov2022enhancement}, and superconducting pairing \cite{cox1989virtual,little2007determination}.\par

Experimentally, $dd$ excitations have been extensively studied by Resonant Inelastic X-ray Scattering (RIXS), which provides a direct probe of both the magnitude and the symmetry of the crystal-field splitting \cite{Ament_RMP}. To date, the most widely used framework for analyzing $dd$ excitations in RIXS spectra is the single-ion model \cite{Ghiringhelli_PRL_2004,Moretti_NJP}. For $d^9$ systems \cite{PhysRevB.99.134517, Rossi_PhysRevB.104.L220505}, this approach takes a particularly simple form, making it a practical alternative to more computationally demanding methods such as \textit{ab initio} calculations \cite{hozoi2011ab} and exact diagonalization \cite{Kuzian_PhysRevB.104.085154}.\par

In recent years, considerable effort has been devoted to the development of machine-learning algorithms for the analysis of diffraction \cite{corriero2023crystalmela}, absorption \cite{Timoshenko_PhysRevLett.120.225502, Rankine2020} and spectroscopy \cite{fischetti2025deep,jung2023automatic,schmid2023deconvolution,fan2019deep} data. In particular, deep neural networks have emerged as powerful tools for the deconvolution of all types of spectra, allowing, in principle, relevant information to be extracted directly from experimental data. This motivates their application to RIXS spectra, since the success in the relatively simple $d^9$ case could open the way to more complex systems, where the $dd$ manifold is harder to analyze.\par


In this Letter, we test the single-ion description of $dd$ excitations in YBa$_2$Cu$_3$O$_6$ and La$_2$CuO$_4$ by analyzing the same set of RIXS spectra with conventional global fitting and a newly developed convolutional neural network trained on single-ion simulations. Since the two methods use the same geometry-dependent $dd$ intensities but extract the spectral parameters in different ways, their comparison provides a stringent consistency test of the model. The outcome of the two analyses agree for YBa$_2$Cu$_3$O$_6$, whereas no unique determination of the orbital energy splittings is obtained for La$_2$CuO$_4$.



\begin{figure}
    \includegraphics[width=\columnwidth]{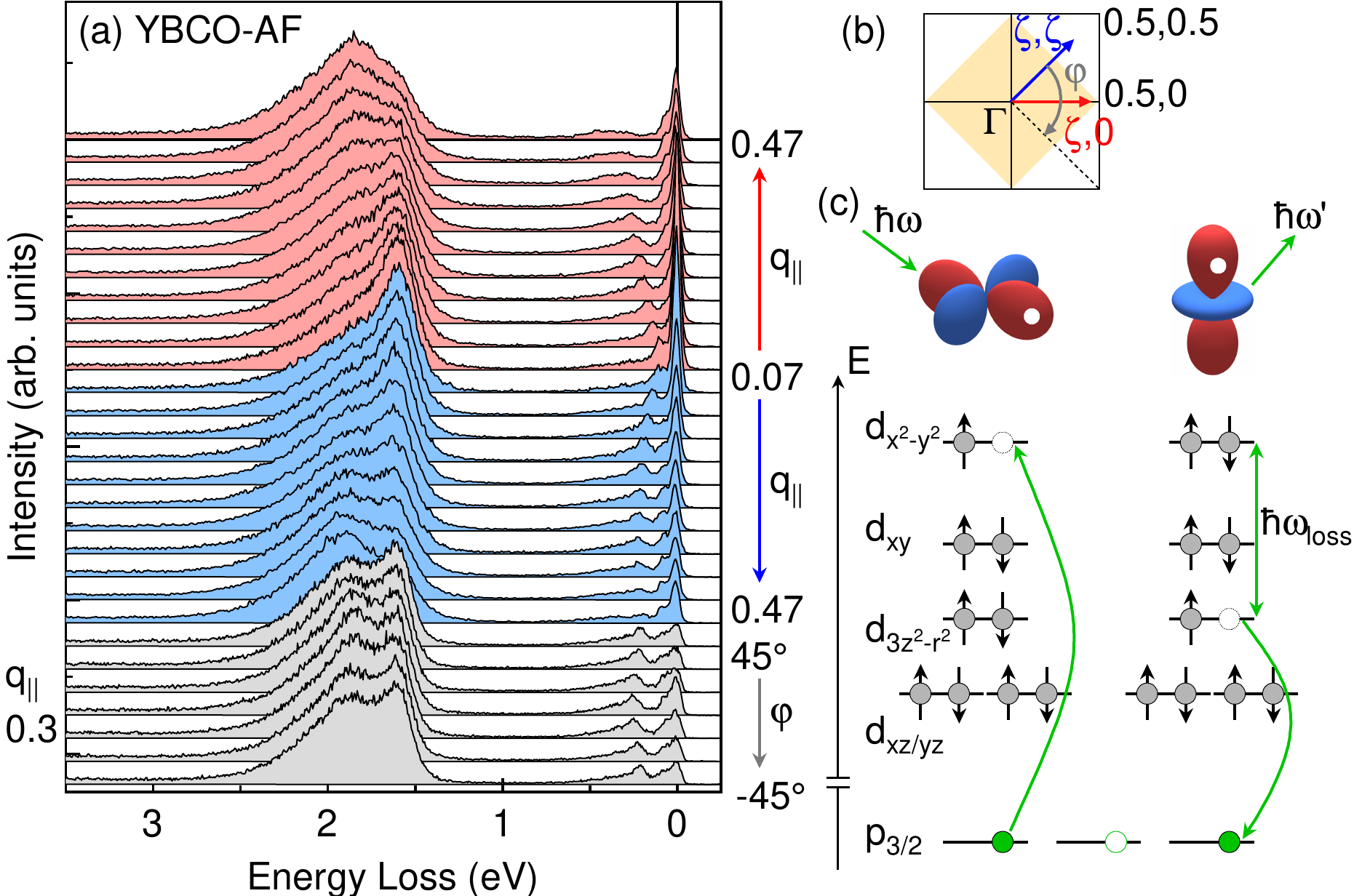}
    \caption{(a) RIXS spectra at the Cu L$_3$ edge for YBCO-AF, with $\sigma$-polarized incident light. The scans along the ($\zeta$,0) and ($\zeta$,$\zeta$) directions were measured on the 50-nm-thick sample, whereas the azimuthal $\varphi$ scan at fixed $\mathrm{q_{\parallel}=0.3}$\,r.l.u. was measured on the 100-nm-thick sample. The corresponding regions of reciprocal space are shown in (b). (c) Schematic of the $dd$ excitation process at the Cu L$_3$ edge, showing the crystal-field splitting of the Cu $3d$ orbitals. Within the single-ion model, the relative $dd$ excitation intensities are determined by the scattering geometry, while the energies are extracted from the spectral analysis.}
    \label{fig:Fig1}
\end{figure}

RIXS spectra of antiferromagnetic YBa$_2$Cu$_3$O$_6$ (YBCO-AF,  thicknesses $t= 50$ and $100\,\mathrm{nm}$) and La$_2$CuO$_4$ (LCO-AF, $t = 30\,\mathrm{nm}$) thin films were measured at the Cu L$_3$ edge on the ID32 beamline of the ESRF\cite{BROOKES_ID32}, with 40--45\,meV energy resolution and $\sigma$-polarized incident light. The in-plane momentum transfer is denoted by $\mathbf{q}_{\parallel}$ and expressed in reciprocal lattice units (r.l.u.). Sample growth, characterization, and data precessing are described in the End Matter. Figure\,\ref{fig:Fig1}\,(a) displays RIXS spectra for the YBCO-AF along different directions in the Brillouin zone, see Panel\,(b). The $dd$ excitations lie in the energy-loss range between 1\,eV and 3\,eV, while a weak charge-transfer (CT) feature appears at higher energy \cite{Ghiringhelli_PRL_2004}. Below 1\,eV, elastic, phonon \cite{Braicovich_PRR,Rossi_PRL,zinouyeva2026influence}, magnon \cite{peng2017influence,Minola_PRL_2015}, and bimagnon \cite{singh2025bimagnon,Chaix_magn_bim} excitations contribute to the spectra. These low-energy peaks, as well as CT, are not discussed in this work.\par 

In a single-ion picture, $dd$ excitations are local orbital excitations within the crystal-field-split Cu $3d$ manifold. In undoped layered cuprates all sites are divalent ($3d^9$ configuration) and the Cu$^{2+}$ ground state is commonly described by a single hole with  $x^{2}-y^{2}$ symmetry \cite{PhysRevLett.68.2543}. In YBCO-AF and LCO-AF, and in most of cuprates, the Cu ion is coordinated in a tetragonally distorted octahedral $C_4$ symmetry if oxygen buckling is neglected. The tetragonal crystal field removes the degeneracy among $x^{2}-y^{2}$, $3z^{2}-r^{2}$, $xy$, and $xz$/$yz$,  which remain unsplit (see Fig.\,\ref{fig:Fig1}\,(c)). Therefore the $dd$ excited states correspond to transferring the $3d$ hole from the $x^{2}-y^{2}$ to one of the other orbitals. \par

Whereas intra-$3d$ transitions are dipole forbidden in optical spectroscopy, they are allowed in L$_3$ RIXS through the double resonant $2p \leftrightarrow 3d$ transitions \cite{Ament_RMP}. Resonant scattering probabilities are calculated within the second-order Kramers--Heisenberg formalism in the electric-dipole approximation, using the dipole operator expressed in the cubic-harmonics basis and accountng for the angular dependence in the two steps of the RIXS process \cite{Moretti_NJP} (see Fig.\,\ref{fig:Fig1}\,(c)). 
The scattering geometry defines also the magnitude and orientation of $\mathbf{q}_{\parallel}$. 

\begin{figure}
    \includegraphics[width=\columnwidth]{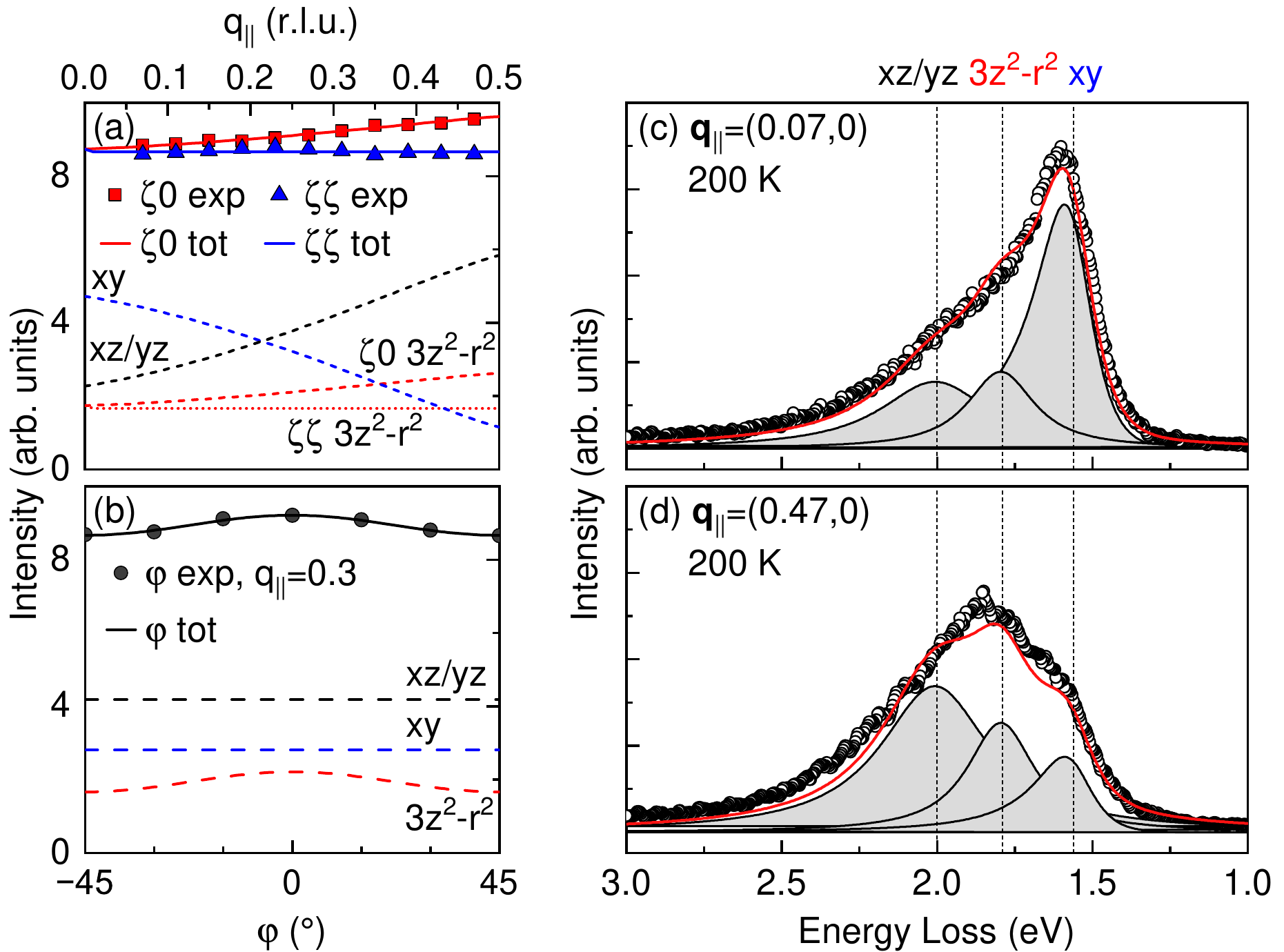}
    \caption{(a) $q_{\parallel}$ and (b) $\varphi$-dependence of the $dd$ excitation cross sections for $\sigma$-incident polarization, calculated within the single-ion model (solid and dashed lines) with markers indicating the energy-integrated intensity of the raw experimental spectra between 1 and 3\,eV. (c), (d) Examples of the $dd$ excitation decomposition for YBCO-AF at two different $\mathbf{q}$ values within the single-ion model. The areas of the fit components are proportional to the single-ion cross sections, while the peak positions (indicated by the vertical dashed lines) and linewidths are the same for all $\mathbf{q}$. We notice that the energy assigned here to $3z^2-r^2$ and $xz/yz$ is different from previous studies \cite{PhysRevB.107.134513,Moretti_NJP, PhysRevB.99.134517}. The basis for this reassignment is the $\varphi$-dependence, as discussed in the text.}
    \label{fig:fig_2}
\end{figure}

Within the single ion model the $dd$ excitations are interpreted as three non-dispersing peaks, labeled $3z^{2}-r^{2}$, $xy$, and $xz$/$yz$ according to the symmetry of the hole in the final state. For each $dd$ excitation, the spin-flip and non-spin-flip channels are assumed to share the same excitation energy because the superexchange interaction among different orbitals is deemed negligible. The energy of each peak is determined by the crystal field strength and the intensity (area) depends on the experimental geometry and can be calculated as explained above \cite{Moretti_NJP}.  Figure\,\ref{fig:fig_2}\,(a,\,b) shows the theoretical $\mathbf{q}$-dependence of the three excitations for 3 paths in the reciprocal space. Their sum is compared to the experimental spectra intensity integrated over the 1-3\,eV energy-loss range, showing good agreement along all investigated directions. This observation validates the use of the single ion model for the interpretation of $dd$ excitations in RIXS spectra of cuprates. We note that only the $3z^2-r^2$ peak intensity is expected to depend on $\varphi$, reaching a maximum at $\varphi = 0^\circ$ and a minimum at $\varphi = 45^\circ$. \par


Our goal is to extract the crystal field energies from the experimental spectra. We first follow Moretti Sala et al. \cite{Moretti_NJP} and apply a global fitting (GF) of the spectra measured with different geometries, see Fig.\,\ref{fig:fig_2}\,(c,\,d). Symmetric Voigt functions are used to simulate the $3z^2-r^2$ and $xz/yz$ peak shapes, while the $xy$ peak is described by a skewed Voigt (asymmetry parameter  $\alpha$) \cite{Azzalini1985,Dockery2025}. The peak intensities are constrained to follow the single-ion cross sections, the Gaussian contribution is fixed by the experimental resolution and the 7 fitting parameters are the energy positions, the Lorentzian widths and asymmetry parameter (when applicable).\par

\begin{figure}
    \includegraphics[width=\columnwidth]{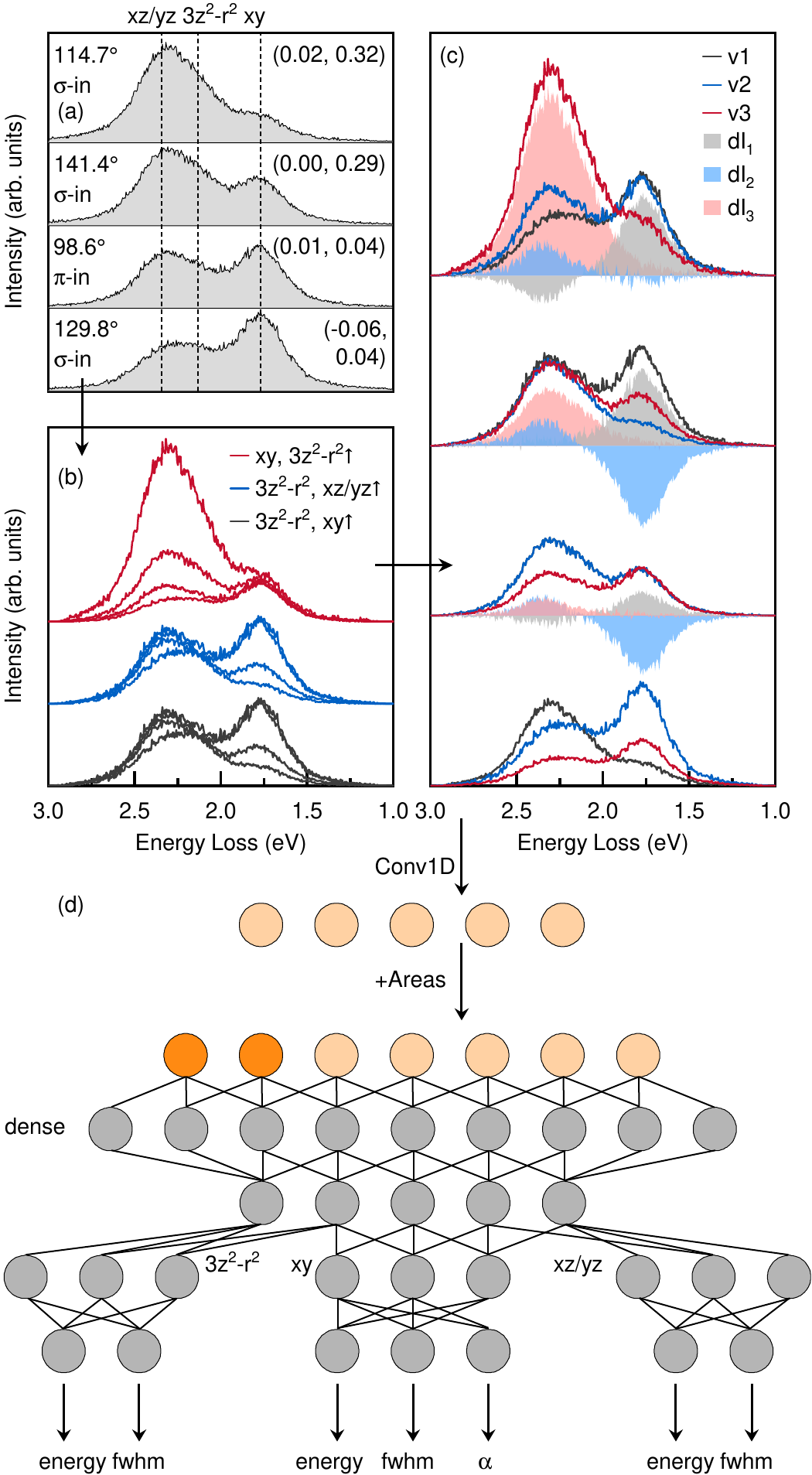}
    \caption{(a) Four simulated spectra corresponding to different experimental geometries, sharing the $dd$ excitation energies (dashed lines), linewidth and $\alpha$ (not shown). 
    (b) Construction of three input variants. In each variant, one selected $dd$ peak is rescaled to a common intensity, and are then reordered according to the increasing area of another peak: fixed $A_{3z^2-r^2}$ with increasing $A_{xy}$ (v1, gray), fixed $A_{3z^2-r^2}$ with increasing $A_{xz/yz}$ (v2, blue), and fixed $A_{xy}$ with increasing $A_{3z^2-r^2}$ (v3, red). (c) Network input construction. For each geometry index  $k=1,\dots,4$, we form one input by taking, for each variant $v=1,2,3$, the spectrum at index $k$, $I_k^{(v)}(E)$, together with its corresponding difference spectrum $I_k^{(v)}(E)-I_1^{(v)}(E)$ (colored areas). These six spectra define a six-channel input. (d) Schematics of the neural network: a shared convolutional encoder extracts spectral features, which are combined with rescaled $dd$ areas and processed by fully connected layers to predict excitation energies, FWHM and $\alpha$ values. }
    \label{fig:fig_ai}
\end{figure}

\begin{figure*}
    \includegraphics[width=\textwidth]{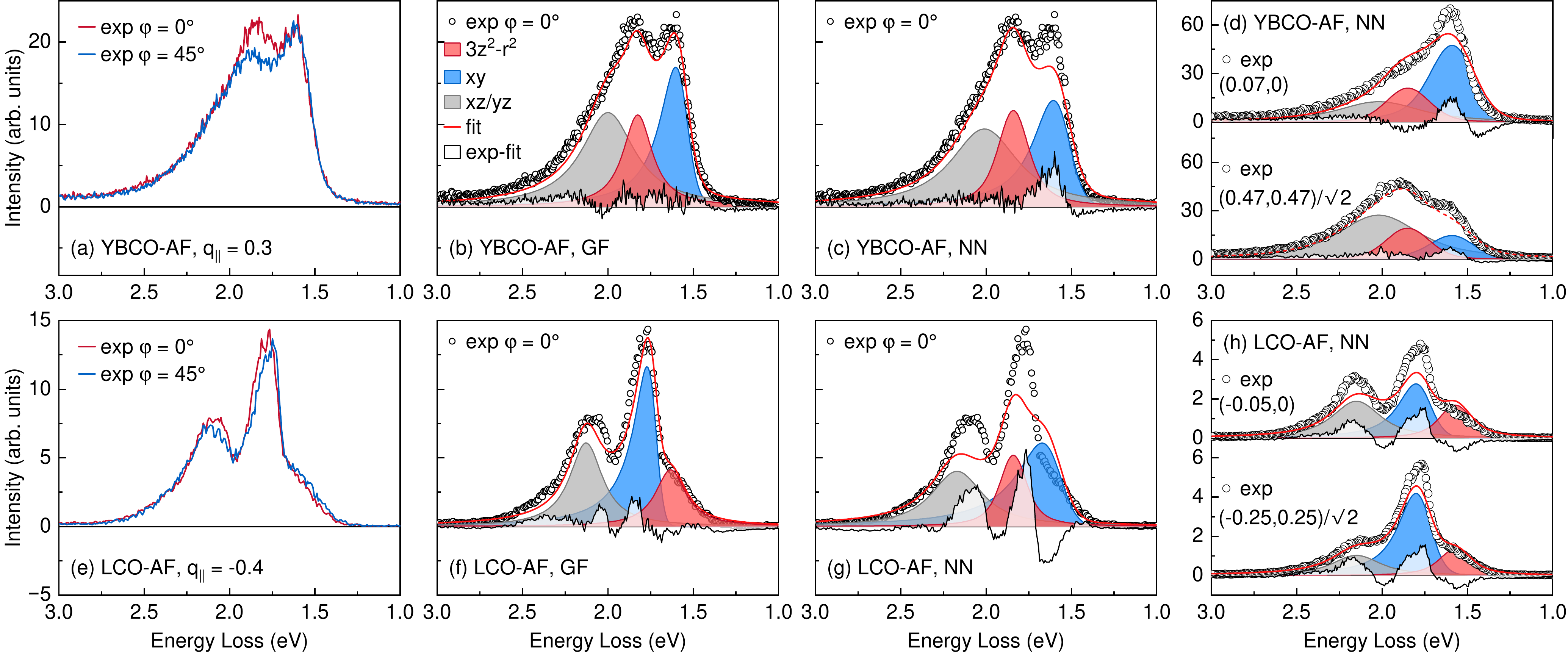}
    \caption{(a) RIXS spectra of YBCO-AF measured at $\varphi=0^{\circ}$ and $45^{\circ}$ with $\sigma$-polarized incident light. (b) Global fit (GF) of a YBCO-AF RIXS spectrum at $\varphi=0^{\circ}$. The black curve with white filling shows the difference between the experimental data and the total fit. (c,d) Neural-network (NN) results for the $\varphi$ scan and $(\zeta,0)$, $(\zeta,\zeta)$ scans. The NN predicts the peak energies, total FWHMs, and $\alpha$. These parameters are used to reconstruct the spectrum within the single-ion line-shape model; a common Gaussian width is adjusted for all peaks, while the Lorentzian width of each peak is fixed by the NN-predicted total FWHMs. (e-h) Same as panels (a-d), but for LCO-AF.}
    \label{fig:fig_results}
\end{figure*}

We compare the experimental data with the theoretical spectra given by the sum of 3 Voigt peaks centered at the $dd$ energies obtained from the fitting. The agreement is very good for several geometries (Fig.\,\ref{fig:fig_2}\,(c,\,d)). However some residual discrepancies are apparent. The agreement could improve by adding fitting parameters (additional peaks, more complex line-shapes), which are however not easy to justify \cite{Martinelli_PRL2024}. Here we want to investigate whether the agreement is not optimal because the global fitting procedure fails to converge to the best values of the parameters, or because the single ion model is partly inadequate in describing $dd$ excitations of undoped cuprates. 
Therefore, we analyze the same data with a different approach, within the same physical model. We use a neural network (NN), trained on a set of artificial spectra based on the single ion model, to estimate the energy and linewidth of the three $dd$ excitations.
Training a neural network requires a large and high-quality dataset for which the target parameters are known. Experimental spectra cannot be used, because the actual energy of $dd$ excitations is not known and the independent spectra available are too few. We generated synthetic spectra using the single ion model over broad ranges of $dd$ energies, scattering geometries and polarizations. Each simulated material is represented by a group of four spectra sharing the same $dd$ energies, FWHM values and asymmetry parameter, but measured in different scattering geometry. To better mimic experimental data, Poisson-statistic noise is added. All details of the data-generation procedure are given in the End Matter.
We trained the neural network on a synthetic dataset containing 300,000 materials, corresponding to a total of $1.2\times10^6$ spectra. Each spectrum is defined over the 1--3\,eV energy-loss range using 400 data points. The model is implemented in TensorFlow/Keras \cite{tensorflow2015-whitepaper} and combines convolutional and fully connected layers \cite{lecun1998gradient,bromley1993signature,szegedy2015going,yu2016multiscale} with ReLU activations \cite{nair2010relu, glorot2011rectifier}. It calculates three excitation energies, the corresponding total FWHM and, only for $xy$, the skewness parameter $\alpha$. Training is performed using a weighted Huber loss \cite{huber1964robust,kendall2018multitask}, which reduces sensitivity to outliers while retaining a quadratic penalty for small errors.\par


The overall NN workflow is illustrated in Fig.\,\ref{fig:fig_ai}. From each group of four spectra we construct three rescaled input variants. In addition, we include difference spectra between geometries, which enhances the intensity dependence on the scattering geometry. They are processed by a shared convolutional encoder and combined with theoretical peak areas to obtain the three excitation energies, the corresponding total FWHM and $\alpha$ values.  Further details are provided in the End Matter and SI.\par

The results of the GF and NN analyses for YBCO-AF are shown in Fig.\,\ref{fig:fig_results}\,(a--d) and summarized in Table\,\ref{tab:summary_all}. The two methods, independently and without imposing constraints on the peak order, lead to the same sequence of increasing energy, namely $xy$, $3z^2-r^2$, and $xz/yz$. The result of GF is shown in Fig.\,\ref{fig:fig_2}\,(c--d). The assignment of an energy $\sim 1.85$\,eV to $3z^2-r^2$ is supported by the joint observation that the $\varphi$ dependence of the experimental spectra is mostly modulated around that energy (Fig.\,\ref{fig:fig_2}\,(b)) and that the theoretical cross sections predict that only the $3z^2-r^2$ intensity depend on the azimuthal angle (Fig.\,\ref{fig:fig_results}\,(a)).
We find the energy of $3z^2-r^2$ smaller than that of $xz/yz$, whereas in earlier studies $3z^2-r^2$ was assigned to the largest energy \cite{Moretti_NJP,PhysRevB.99.134517}. The refinement of the energy of this state has been made possible by higher energy resolution and the use  of a broader set of geometries, including those at $\varphi=45$\,deg \cite{Ghiringhelli_PRL_2004,Moretti_NJP}. 
A discrepancy is observed between the FWHM and $\alpha$ values obtained from the fitting procedure and the values predicted by the neural network. This may be related to the fact that, when the Gaussian component is left free during the fitting, it converges to values about 2--4 times larger than expected from the experimental resolution. The $dd$ line shapes therefore appear to contain additional broadening contributions, possibly arising from phonons \cite{Martinelli_PRL2024, Lee_PhysRevB.89.041104} or other excitations. This extra broadening does not seem to impact on the $dd$ energies but signals the fact that the single ion model cannot account completely for the spectral shape of $dd$ excitations in RIXS.\par



\begin{table*}[t]
\caption{
Summary of peak energies (in eV), asymmetry parameter $\alpha$, widths (in eV), and peak-normalized fit error, quantified by Normalized Root Mean Square Error (NRMSE, in \%), for YBCO-AF and LCO-AF. Here, pol. denotes polarimeter measurements, see SI. The NRMSE is defined as
$\mathrm{NRMSE}_{\mathrm{peak}} = 100 \times \left\langle \sqrt{\frac{1}{N_k}\sum_i (y_{k,i}-y^{\mathrm{fit}}_{k,i})^2}/\max_i |y_{k,i}| \right\rangle_k$,
where $k$ labels the spectra and $i$ the energy points. For LCO-AF, the NRMSE values indicate that the main issue is not simply the residual fit quality, but the lack of a unique and consistent orbital assignment across scans and between GF and NN analyses. For the global fitting, the error bars are on the order of a few meV and are therefore omitted. }
\label{tab:summary_all}
\centering
\small
\setlength{\tabcolsep}{4pt}
\renewcommand{\arraystretch}{1.15}
\begin{tabular*}{\textwidth}{@{\extracolsep{\fill}} llcccccccc @{}}
\toprule
Sample & Scan
& \multicolumn{1}{c}{$E_{3z^2-r^2}$}
& \multicolumn{1}{c}{$E_{xy}$}
& \multicolumn{1}{c}{$E_{xz/yz}$}
& \multicolumn{1}{c}{$\alpha$}
& \multicolumn{1}{c}{$W_{3z^2-r^2}$}
& \multicolumn{1}{c}{$W_{xy}$}
& \multicolumn{1}{c}{$W_{xz/yz}$}
& \multicolumn{1}{c}{NRMSE \%} \\
\cmidrule(lr){3-10}
& 
& GF, NN
& GF, NN
& GF, NN
& GF, NN
& GF, NN
& GF, NN
& GF, NN
& GF, NN\\
\midrule
YBCO-AF & $\varphi$
& 1.83, 1.84
& 1.56, 1.54
& 2.00, 2.01
& 0.20, 0.49
& 0.21, 0.22
& 0.20, 0.26
& 0.38, 0.47
& 3.89, 6.06\\

YBCO-AF & $(\zeta,0)$, $(\zeta,\zeta)$
& 1.79, 1.85
& 1.56, 1.54
& 2.00, 2.02
& 0.12, 0.45
& 0.26, 0.30
& 0.23, 0.31
& 0.42, 0.57
& 3.69, 5.57\\

YBCO-AF & pol.
& 1.84, 1.88
& 1.59, 1.58
& 2.08, 2.04
& 0.05, 0.24
& 0.20, 0.25
& 0.21, 0.23
& 0.48, 0.53
& 7.15, 7.51 \\
\midrule
LCO-AF & $\varphi$
& 1.63, 1.84
& 1.72, 1.58
& 2.13, 2.17
& 0.41, 0.49
& 0.21, 0.19 
& 0.17, 0.32
& 0.23, 0.36
& 3.35, 9.40\\

LCO-AF & $(\zeta,0)$, $(\zeta,\zeta)$
& 1.63, 1.57
& 1.75, 1.75
& 2.15, 2.15
& 0.28, 0.48
& 0.23, 0.23
& 0.17, 0.23
& 0.25, 0.34
& 3.59, 6.72\\
\bottomrule
\end{tabular*}
\end{table*}


In the case of LCO-AF, the same analysis reveals discrepancies, especially when we consider the $\varphi$ scan, see Fig.\,\ref{fig:fig_results}\,(e). According to the literature \cite{Moretti_NJP,peng2017influence, ivashko2019strain} and the global single-ion fit, see Fig.\,\ref{fig:fig_results}\,(f), the expected order of the $dd$ excitations is $3z^2-r^2$, $xy$, and $xz/yz$. However, this sequence is not totally compatible with our more complete set of experimental data. In particular, the peak around 1.5\,eV shows a slightly higher intensity at $\varphi = 45^\circ$ than at $\varphi = 0^\circ$, whereas the characteristic angular dependence of the $3z^2-r^2$ excitation is observed near 2\,eV, see Fig.\,\ref{fig:fig_results}\,(e).  Consistently, the neural network, which is trained on the differences between the spectra, assigns the $3z^2-r^2$ peak to 1.84\,eV, while the positions of the $xy$ and $xz/yz$ peaks are selected in the energy region where the spectra in the $\varphi$ scans overlap, see Fig.\,\ref{fig:fig_results}\,(g). 
We further note that in the global fitting, the $3z^2 - r^2$ peak is found at 1.85\,eV when all peaks are modeled with symmetric Voigt profiles, see SI. However, when the neural network is applied to the $(\zeta,0)$ and $(\zeta,\zeta)$ scans, it predominantly assigns the $3z^2-r^2$ excitation to $\sim 1.6$\,eV, in contrast to the $\varphi$ scan results, see Fig.\,\ref{fig:fig_results}\,(h) and SI. This discrepancy may indicate the presence of momentum-dependent spectral contributions beyond the current single-ion training model. Such contributions could modify the apparent line shape differently along different directions in the Brillouin zone, making the neural network predictions sensitive to the particular set of four spectra used as input.\par

The origin of the discrepancy between the single-ion description and the experimental spectra in LCO-AF remains unclear. Deviations from the single-ion picture have been previously reported \cite{Robarts_dd_in_LCO}, suggesting that additional interactions beyond a purely local model are relevant. Proposed mechanisms include coupling to phonons \cite{Martinelli_PRL2024, Lee_PhysRevB.89.041104} and/or magnons \cite{Kuzian_PhysRevB.104.085154,Comin_PhysRevB.99.045105, Barantani_PhysRevX.12.021068}, the role of Hund’s exchange \cite{singh2020stability}, and hybridization effects between $x^2-y^2$ and $3z^2-r^2$ orbitals \cite{matt2018direct,Sakakibara_2010,Sakakibara_2012} and with other low-lying electronic states. A systematic polarization-resolved RIXS study might help clarify which of these contributions can be relevant in LCO-AF spectra. These observations point to the need for a more complete theoretical framework to account for the limitations of the single-ion description in LCO-AF, reconcile its validity in YBCO-AF, and explain similar anomalies reported in CaCuO$_2$ \cite{Moretti_NJP, Martinelli_PRL2024} and Nd$_2$CuO$_4$ \cite{Comin_PhysRevB.99.045105}.
Finally, we note that our analysis is performed on antiferromagnetic undoped cuprates, where the $dd$ excitations are relatively sharp. Upon doping, the $dd$ peaks show substantial broadening and strong overlap \cite{PhysRevB.99.134517}. In addition, in the 1--3\,eV region there is a strong electron-hole continuum contribution, that cannot be accounted for by the single-ion model \cite{PhysRevB.99.134517,Minola_PRL_2015}. Therefore, a neural network for RIXS spectra of doped cuprates should include the electron-hole continuum contributions in addition to the localized orbital excitations investigated in this Letter.\par

Our results can be summarized as follows. For YBCO-AF, a conventional single-ion fit already provides a consistent description of the $dd$ excitation manifold, as confirmed by its agreement with the neural network analysis. By contrast, in LCO-AF the fact that the two approaches cannot converge to a unique solution highlights the limitations of the underlying single-ion model itself. In this work, we show that the machine learning analysis both provides a valid alternative to GF for the parameter extraction from experimental spectra and offers a stringent test of the validity of the physical framework on which it is trained. In conclusion the comparison of the two methods stimulates the development of more complete theoretical frameworks for $dd$ excitation spectra of cuprates beyond the single-ion model. 
Moreover, the approach proposed here for the simple case of cuprates can  be extended to the analysis and interpretation of RIXS spectra of other correlated materials, namely $3d^{2-8}$ systems, where single-ion calculations are more complex and require a larger number of parameters, whose optimization is commonly made by expert human analysis possibly supported by optimization algorithms \cite{tnqm-ttj3}. In such cases, training on simulated spectra (e.g., 2D maps generated with \textsc{Quanty} \cite{Haverkort_2016} or EDRIXS \cite{wang2019edrixs}) could enable the straightforward extraction of crystal-field and atomic parameters for materials of interest.\par


\begin{acknowledgments}
We are deeply grateful to Guido Caldarelli and Giulia Fischetti for useful discussions. We thank Matteo Rossi and Lucio Braicovich for their contribution to the RIXS measurements. The experimental data were collected at the beamline ID32 of the ESRF during experiments HC4000, HC4149, HC4171, HC4412 \cite{HC4412}, HC5221 \cite{HC5221}; we thank the whole technical and scientific staff of ID32 for their excellent support.  Myfab is acknowledged for support and for access to the nanofabrication laboratories at Chalmers, where the YBCO films were grown. The authors gratefully acknowledge the Data Center ReCaS-Cosenza for access to the high-performance computing infrastructure used to train the neural network and for technical support. M.Z., L.M., D.D.C., G.M., F.R., M.M.S., and G.G. acknowledge support by the Italian Ministry of University and Research (MUR) through the project PRIN2020 ``QT-FLUO'' ID 20207ZXT4Z of the Ministry for University and Research (MUR) of Italy. M.Z., G.M., F.R. and G.G. acknowledge support  by the Italian Ministry of University and Research (MUR) through the project FIS2 ``abovEF'' FIS-2023-02406. F.L. acknowledges support by the Swedish Research Council (VR) under the Project 2022-04334 and by the Knut and Alice Wallenberg foundation under the Projects KAW 2023-0341 and KAW 2024-0129. 
\end{acknowledgments}

\section{\label{sec:level1}End Matter}
\paragraph*{Sample details and RIXS measurements.} Thin films of YBa$_2$Cu$_3$O$_6$ (hole doping $p < 0.04$, thicknesses $t= 50$ and $100\,\mathrm{nm}$, lattice parameters $a = b = 3.87\,\mathrm{\AA}$, $c = 11.83\,\mathrm{\AA}$) and La$_2$CuO$_4$ ($p \sim 0$, $t = 30\,\mathrm{nm}$, $a = b = 3.76\,\mathrm{\AA}$, $c = 13.15\,\mathrm{\AA}$) were grown by pulsed laser deposition on SrTiO$_3$ and SrLaAlO$_4$ substrates, respectively, following the procedure described in Ref.\,\cite{arpaia_2018_YBCO,Martinelli_PRL2024}. RIXS measurements were performed at the ID32 beamline of the European Synchrotron Radiation Facility \cite{BROOKES_ID32}, with the energy resolution of 40-45\,meV at the Cu L$_3$ edge ($\approx931$\,eV). The in-plane momentum transfer is indicated with $\mathbf{q_{\parallel}}$, here in reciprocal lattice units (r.l.u.). We employed $\sigma$ polarization (perpendicular to the scattering plane), and the scattered beam was recorded either with or without polarization analysis \cite{PhysRevB.99.134517}. All RIXS spectra were normalized to the incident photon flux, and corrected for self-absorption effects, which are however almost negligible for $dd$ excitation \cite{arpaia2023signature}.
\paragraph*{Synthetic-data generation and neural-network architecture.} We summarize the synthetic-data generation, the preprocessing steps, and the neural-network architecture used in this work. Each spectrum is modeled as the sum of three Voigt peaks: symmetric Voigt for $3z^2-r^2$ and $xz/yz$ peaks, and skew-Voigt for $xy$ peak. The peak energies are randomly sampled between 1.3 and 2.5\,eV for the $3z^2-r^2$ and $xz/yz$ orbitals and between 1.3 and 2.0\,eV for the $xy$ orbital. The Lorentzian HWHM and Gaussian FWHM are randomly selected within the ranges 0.03--0.25\,eV and 0.025--0.3\,eV, respectively. The skewness parameter $\alpha$ is arbitrarily selected between 0 and 0.5. The incident polarization is randomly chosen between $\sigma$ and $\pi$, and the spectra correspond to the sum of the outgoing $\sigma$ and $\pi$ channels. The experimental geometries are randomly generated within the kinematically accessible range of the RIXS setup. For each material, four geometries are sampled by varying the scattering angle $2\theta$, the in-plane azimuthal angle $\varphi$, and the magnitude of the in-plane momentum transfer $|\textbf{q}_{\parallel}|$. The scattering angle is restricted to $60^\circ \leq 2\theta \leq 154^\circ$. For a given value of $2\theta$, the maximum accessible momentum transfer is determined by the photon wavelength according to
\[
|q|_{\max} =
\frac{4\pi}{\lambda}
\sin\left(\frac{2\theta}{2}\right)
\frac{a}{2\pi},
\]
where $\lambda$ is the photon wavelength and $a$ is the in-plane lattice parameter. The factor $a/(2\pi)$ converts the momentum transfer from absolute units, \AA$^{-1}$, to reciprocal-lattice units. The value of $|\textbf{q}_{\parallel}|$ is then randomly sampled between zero and this maximum allowed value. To cover different experimental acquisition strategies, several geometry families are generated, including $\varphi$ scans at fixed $2\theta$ and $|\textbf{q}_{\parallel}|$, $|\textbf{q}_{\parallel}|$ scans at fixed $2\theta$ and $\varphi$, scans involving two azimuthal angles, scans in $2\theta$, and fully random geometries.\par
Poisson noise is added on the wide uniform energy grid by first converting the noiseless intensity value $I_i$ at each energy bin into an expected number of photon counts,
\[
\lambda_i = I_i \, \Delta E \, N_{\mathrm{counts}}^{\mathrm{eff}},
\]
where $\Delta E$ is the energy-bin width and $N_{\mathrm{counts}}^{\mathrm{eff}}$ sets the overall counting statistics. The observed counts are then sampled as $n_i \sim \mathrm{Poisson}(\lambda_i)$ and converted back to an intensity, $I_i^{\mathrm{noisy}} = n_i/(N_{\mathrm{counts}}^{\mathrm{eff}}\Delta E)$, before cutting and interpolating the spectra onto the final energy grid. In some cases, this procedure produces simulated spectra that are noisier than the experimental spectra, which makes the training set conservative with respect to counting noise.\par

Before training, each set of four spectra is transformed into three complementary input variants. In each variant, a selected $dd$ peak is rescaled to have the same area in all four geometries, and the same scaling factors are applied to the corresponding target areas. In the first, second, and third variants, the reference peak is chosen as $3z^2-r^2$, $3z^2-r^2$, and $xy$, respectively. After rescaling, the spectra are ordered according to area of another $dd$ peak: by increasing $A_{xy}$ in the first variant, by increasing $A_{xz/yz}$ in the second, and by increasing $A_{3z^2-r^2}$ in the third. Any remaining degeneracies are resolved using a secondary area criterion, namely increasing $A_{xz/yz}$, $A_{xy}$, and $A_{xz/yz}$ for the first, second, and third variants, respectively, and, if necessary, the original geometry index. The four spectra in each variant are then jointly normalized so that the sum of their integrated intensities is unity. This procedure removes trivial scale factors while preserving the relative intensity differences that encode the geometry dependence of the cross section.\par

For each variant, we also construct difference spectra $I_k(E)-I_1(E)$, which emphasize geometry-dependent intensity contrasts. The network input is then built by grouping spectra with the same geometry index $k$ across the three variants. For each $k=1,\dots,4$, we combine: (i) the three spectra from the three variants at geometry $k$, and (ii) the corresponding difference spectra $I_k(E)-I_1(E)$ for each variant. This procedure defines four groups per sample each containing six spectra.\par

A shared multiscale convolutional encoder \cite{lecun1998gradient,bromley1993signature,szegedy2015going,yu2016multiscale} is applied independently to the four groups. The encoder consists of parallel convolutional branches with different kernel sizes and dilation rates, enabling the simultaneous capture of narrow spectral structures and broader line-shape variations. To prevent an excessive reliance on the contrast channels, channel-wise dropout is applied during training to the three difference-spectrum channels.\par

In parallel, the model receives a second input containing the rescaled and normalized $dd$ areas. In each preprocessing variant, the area values are sorted using the same ordering applied to the spectra, such that the two inputs remain aligned. For each of the four ordered positions, the area input contains nine values: the three orbital areas from each of the three variants. These values are processed by dense layers with weights shared across the four ordered positions and then concatenated with the corresponding spectral features. The four combined representations are finally flattened and processed by additional dense layers, allowing the network to learn correlations among the ordered spectral groups.\par

In the final stage, the network splits into three parallel prediction heads, one for each $dd$ excitation. Each head outputs the excitation energy, the corresponding total FWHM and,  only for $xy$, $\alpha$ parameter. Training is performed using a weighted Huber loss \cite{huber1964robust,kendall2018multitask},
\begin{equation}
\mathcal{L}_{\delta}(y,\hat{y}) =
\begin{cases}
\dfrac{1}{2}(y-\hat{y})^2 & \text{if } |y-\hat{y}| \le \delta \\
\delta \left( |y-\hat{y}| - \dfrac{1}{2}\delta \right) & \text{otherwise},
\end{cases}
\end{equation}
where $y$ and $\hat{y}$ denote the true and predicted values, respectively, and $\delta$ sets the transition between quadratic and linear regimes. We use $\delta=0.2$, applied separately to the energy, linewidth and $\alpha$ outputs.\par

For the experimental data, the input representation follows the same procedure as used for the synthetic dataset. Four measured spectra are interpolated onto the training energy-loss grid, and the $dd$ areas are computed from the scattering geometry and polarization within the single-ion formalism. These cross-sections are used in the three preprocessing variants used in training: one orbital channel is selected to have equal integrated intensity across the four geometries, the spectral order is determined by the rescaled $dd$ areas, and the four spectra are normalized jointly to unit total intensity. For each variant, spectra differences with respect to the first geometry are calculated, producing the same six-channel input representation used in training: three spectra and three difference spectra. The trained model then predicts the three excitation energies, the three total FWHM values, and the $xy$ asymmetry parameter $\alpha$. In this way, the analysis of experimental data remains fully consistent with the synthetic training protocol.

\paragraph*{Role of the neural-network analysis.} The role of neural network is to perform a different test of the same model. A global fit can adapt to the data even when the physical model is not fully correct, since changes in energies, linewidths, and asymmetries may produce a curve that follows the experimental spectrum reasonably well. However, a good fit curve does not guarantee that the physical assignment of the peaks is unique or correct. The neural network performs a complementary check. During training, it only sees spectra generated within the single-ion model, and therefore learns which spectral shapes, geometry-dependent variations, and parameter combinations are compatible with that model. When applied to experimental spectra, the NN gives a robust assignment only if those spectra resemble the single-ion spectra used for training. For YBCO-AF, GF and NN give the same assignment, indicating that the spectra belong consistently to the family of spectra described by the single-ion model. For LCO-AF, the GF finds acceptable solutions, but the NN does not identify a unique and stable assignment. In particular, the NN assignment changes depending on the set of spectra used as input. The issue is therefore not only that the fit is imperfect, but that the experimental data cannot be mapped uniquely onto the family of single-ion spectra.

\bibliography{ddbiblio}

@article{Ament_RMP,
  title = {{Resonant inelastic x-ray scattering studies of elementary excitations}},
  author = {Ament, Luuk J. P. and van Veenendaal, Michel and Devereaux, Thomas P. and Hill, John P. and van den Brink, Jeroen},
  journal = {Rev. Mod. Phys.},
  volume = {83},
  issue = {2},
  pages = {705--767},
  numpages = {0},
  year = {2011},
  month = {Jun},
  publisher = {American Physical Society},
  doi = {10.1103/RevModPhys.83.705},
  url = {https://link.aps.org/doi/10.1103/RevModPhys.83.705}
}

@article{arpaia_2018_YBCO,
  title={{Probing the phase diagram of cuprates with \ch{YBa_2Cu_3O_{7- $\delta$}} thin films and nanowires}},
  author={Arpaia, Riccardo and Andersson, Eric and Trabaldo, Edoardo and Bauch, Thilo and Lombardi, Floriana},
  journal={Physical Review Materials},
  volume={2},
  number={2},
  pages={024804},
  year={2018},
  publisher={APS},
  url = {https://journals.aps.org/prmaterials/abstract/10.1103/PhysRevMaterials.2.024804}
}

@article{Martinelli_PRL2024,
  title = {{Collective Nature of Orbital Excitations in Layered Cuprates in the Absence of Apical Oxygens}},
  author = {Martinelli, Leonardo and Wohlfeld, Krzysztof and Pelliciari, Jonathan and Arpaia, Riccardo and Brookes, Nicholas B. and Di Castro, Daniele and Fernandez, Mirian G. and Kang, Mingu and Krockenberger, Yoshiharu and Kummer, Kurt and McNally, Daniel E. and Paris, Eugenio and Schmitt, Thorsten and Yamamoto, Hideki and Walters, Andrew and Zhou, Ke-Jin and Braicovich, Lucio and Comin, Riccardo and Sala, Marco Moretti and Devereaux, Thomas P. and Daghofer, Maria and Ghiringhelli, Giacomo},
  journal = {Phys. Rev. Lett.},
  volume = {132},
  issue = {6},
  pages = {066004},
  numpages = {8},
  year = {2024},
  month = {Feb},
  publisher = {American Physical Society},
  doi = {10.1103/PhysRevLett.132.066004},
  url = {https://link.aps.org/doi/10.1103/PhysRevLett.132.066004}
}

@article{BROOKES_ID32,
title={{The beamline ID32 at the ESRF for soft X-ray high energy resolution resonant inelastic X-ray scattering and polarisation dependent X-ray absorption spectroscopy}},
  author={Brookes, N. B. and Yakhou-Harris, F and Kummer, K and Fondacaro, A and Cezar, JC and Betto, D and Velez-Fort, E and Amorese, A and Ghiringhelli, G and Braicovich, L and others},
  journal={Nuclear Instruments and Methods in Physics Research Section A: Accelerators, Spectrometers, Detectors and Associated Equipment},
  volume={903},
  pages={175--192},
  year={2018},
  publisher={Elsevier},
doi = {https://doi.org/10.1016/j.nima.2018.07.001},
url = {https://www.sciencedirect.com/science/article/pii/S0168900218308234}
}

@article{Braicovich_PRR,
  title = {{Determining the electron-phonon coupling in superconducting cuprates by resonant inelastic x-ray scattering: Methods and results on \ch{Nd_{1+x}Ba_{2-x}Cu_{3}O_{7-$\delta$}}}},
  author = {Braicovich, Lucio and Rossi, Matteo and Fumagalli, Roberto and Peng, Yingying and Wang, Yan and Arpaia, Riccardo and Betto, Davide and De Luca, Gabriella M. and Di Castro, Daniele and Kummer, Kurt and Moretti Sala, Marco and Pagetti, Mattia and Balestrino, Giuseppe and Brookes, Nicholas B. and Salluzzo, Marco and Johnston, Steven and van den Brink, Jeroen and Ghiringhelli, Giacomo},
  journal = {Phys. Rev. Res.},
  volume = {2},
  issue = {2},
  pages = {023231},
  numpages = {19},
  year = {2020},
  publisher = {American Physical Society},
  doi = {10.1103/PhysRevResearch.2.023231},
  url = {https://link.aps.org/doi/10.1103/PhysRevResearch.2.023231}
}

@article{Rossi_PRL,
  title = {{Experimental Determination of Momentum-Resolved Electron-Phonon Coupling}},
  author = {Rossi, Matteo and Arpaia, Riccardo and Fumagalli, Roberto and Moretti Sala, Marco and Betto, Davide and Kummer, Kurt and De Luca, Gabriella M. and van den Brink, Jeroen and Salluzzo, Marco and Brookes, Nicholas B. and Braicovich, Lucio and Ghiringhelli, Giacomo},
  journal = {Phys. Rev. Lett.},
  year = {2019},
  volume = {123},
  issue = {2},
  pages = {027001},
  numpages = {6},
  month = {Jul},
  publisher = {American Physical Society},
  doi = {10.1103/PhysRevLett.123.027001},
  url = {https://link.aps.org/doi/10.1103/PhysRevLett.123.027001}
}

@article{peng2017influence,
  title={Influence of apical oxygen on the extent of in-plane exchange interaction in cuprate superconductors},
  author={Peng, YY and Dellea, G and Minola, M and Conni, M and Amorese, A and Di Castro, D and De Luca, GM and Kummer, K and Salluzzo, M and Sun, X and others},
  journal={Nature Physics},
  volume={13},
  number={12},
  pages={1201--1206},
  year={2017},
  publisher={Nature Publishing Group UK London},
  doi = {10.1038/nphys4248}
}

@article{Minola_PRL_2015,
  title = {{Collective Nature of Spin Excitations in Superconducting Cuprates Probed by Resonant Inelastic X-Ray Scattering}},
  author = {Minola, M. and Dellea, G. and Gretarsson, H. and Peng, Y. Y. and Lu, Y. and Porras, J. and Loew, T. and Yakhou, F. and Brookes, N. B. and Huang, Y. B. and Pelliciari, J. and Schmitt, T. and Ghiringhelli, G. and Keimer, B. and Braicovich, L. and Le Tacon, M.},
  journal = {Phys. Rev. Lett.},
  volume = {114},
  issue = {21},
  pages = {217003},
  numpages = {6},
  year = {2015},
  month = {May},
  publisher = {American Physical Society},
  doi = {10.1103/PhysRevLett.114.217003},
  url = {https://link.aps.org/doi/10.1103/PhysRevLett.114.217003}
}

@article{singh2025bimagnon,
  title={{Bimagnon dispersion of La$_2$CuO$_4$ probed by resonant inelastic X-ray scattering}},
  author={Singh, A and Huang, HY and Tsutsui, K and Tohyama, T and Komiya, S and Okamoto, J and Chen, CT and Fujimori, A and Huang, DJ},
  journal={Scientific reports},
  volume={15},
  number={1},
  pages={34183},
  year={2025},
  publisher={Nature Publishing Group UK London},
  doi = {10.1038/s41598-025-15435-5}
}

@article{zinouyeva2026influence,
  title={{The influence of phonon symmetry and electronic structure on the electron-phonon coupling momentum dependence in cuprates}},
  author={Zinouyeva, Maryia and Heid, Rolf and Merzoni, Giacomo and Arpaia, Riccardo and Andreev, Nikolai and Biagi, Marco and Brookes, Nicholas B and Di Castro, Daniele and Kalaboukhov, Alexei and Kummer, Kurt and others},
  journal={npj Quantum Materials},
  year      = {2026},
  volume    = {11},
  pages     = {30},
  publisher={Nature Publishing Group UK London},
  doi = {10.1038/s41535-026-00863-x}
}

@article{Chaix_magn_bim,
  title = {{Resonant inelastic x-ray scattering studies of magnons and bimagnons in the lightly doped cuprate ${\mathrm{La}}_{2\ensuremath{-}x}{\mathrm{Sr}}_{x}{\mathrm{CuO}}_{4}$}},
  author = {Chaix, L. and Huang, E. W. and Gerber, S. and Lu, X. and Jia, C. and Huang, Y. and McNally, D. E. and Wang, Y. and Vernay, F. H. and Keren, A. and Shi, M. and Moritz, B. and Shen, Z.-X. and Schmitt, T. and Devereaux, T. P. and Lee, W.-S.},
  journal = {Phys. Rev. B},
  volume = {97},
  issue = {15},
  pages = {155144},
  numpages = {8},
  year = {2018},
  month = {Apr},
  publisher = {American Physical Society},
  doi = {10.1103/PhysRevB.97.155144},
  url = {https://link.aps.org/doi/10.1103/PhysRevB.97.155144}
}

@article{Ghiringhelli_PRL_2004,
  title = {{Low Energy Electronic Excitations in the Layered Cuprates Studied by Copper ${L}_{3}$ Resonant Inelastic X-Ray Scattering}},
  author = {Ghiringhelli, G. and Brookes, N. B. and Annese, E. and Berger, H. and Dallera, C. and Grioni, M. and Perfetti, L. and Tagliaferri, A. and Braicovich, L.},
  journal = {Phys. Rev. Lett.},
  volume = {92},
  issue = {11},
  pages = {117406},
  numpages = {4},
  year = {2004},
  month = {Mar},
  publisher = {American Physical Society},
  doi = {10.1103/PhysRevLett.92.117406},
  url = {https://link.aps.org/doi/10.1103/PhysRevLett.92.117406}
}

@article{Moretti_NJP,
  title={{Energy and symmetry of dd excitations in undoped layered cuprates measured by Cu $L_3$ resonant inelastic x-ray scattering}},
  author={Sala, M Moretti and Bisogni, Valentina and Aruta, C and Balestrino, G and Berger, H and Brookes, NB and De Luca, GM and Di Castro, D and Grioni, M and Guarise, M and others},
  journal={New Journal of Physics},
  volume={13},
  number={4},
  pages={043026},
  year={2011},
  publisher={IOP Publishing},
  doi = {10.1088/1367-2630/13/4/043026}
}

@article{PhysRevB.99.134517,
  title = {{Polarization-resolved Cu ${L}_{3}$-edge resonant inelastic x-ray scattering of orbital and spin excitations in ${\mathrm{NdBa}}_{2}{\mathrm{Cu}}_{3}{\mathrm{O}}_{7\ensuremath{-}\ensuremath{\delta}}$}},
  author = {Fumagalli, R. and Braicovich, L. and Minola, M. and Peng, Y. Y. and Kummer, K. and Betto, D. and Rossi, M. and Lefran\ifmmode \mbox{\c{c}}\else \c{c}\fi{}ois, E. and Morawe, C. and Salluzzo, M. and Suzuki, H. and Yakhou, F. and Le Tacon, M. and Keimer, B. and Brookes, N. B. and Sala, M. Moretti and Ghiringhelli, G.},
  journal = {Phys. Rev. B},
  volume = {99},
  issue = {13},
  pages = {134517},
  numpages = {12},
  year = {2019},
  month = {Apr},
  publisher = {American Physical Society},
  doi = {10.1103/PhysRevB.99.134517},
  url = {https://link.aps.org/doi/10.1103/PhysRevB.99.134517}
}

@article{Haverkort_2016,
doi = {10.1088/1742-6596/712/1/012001},
url = {https://doi.org/10.1088/1742-6596/712/1/012001},
year = {2016},
month = {may},
publisher = {IOP Publishing},
volume = {712},
number = {1},
pages = {012001},
author = {Haverkort, Maurits W.},
title = {{Quanty for core level spectroscopy - excitons, resonances and band excitations in time and frequency domain}},
journal = {Journal of Physics: Conference Series}
}

@article{wang2019edrixs,
  title={EDRIXS: An open source toolkit for simulating spectra of resonant inelastic x-ray scattering},
  author={Wang, YL and Fabbris, Gilberto and Dean, Mark PM and Kotliar, Gabriel},
  journal={Computer Physics Communications},
  volume={243},
  pages={151--165},
  year={2019},
  publisher={Elsevier},
  doi = {10.1016/j.cpc.2019.04.018}
}

@article{Robarts_dd_in_LCO,
  title = {{Dynamical spin susceptibility in ${\mathrm{La}}_{2}{\mathrm{CuO}}_{4}$ studied by resonant inelastic x-ray scattering}},
  author = {Robarts, H. C. and Garc\'{\i}a-Fern\'andez, M. and Li, J. and Nag, A. and Walters, A. C. and Headings, N. E. and Hayden, S. M. and Zhou, Ke-Jin},
  journal = {Phys. Rev. B},
  volume = {103},
  issue = {22},
  pages = {224427},
  numpages = {11},
  year = {2021},
  month = {Jun},
  publisher = {American Physical Society},
  doi = {10.1103/PhysRevB.103.224427},
  url = {https://link.aps.org/doi/10.1103/PhysRevB.103.224427}
}

@article{singh2020stability,
  title={The stability of hole-doped antiferromagnetic state in a two-orbital model},
  author={Singh, Dheeraj Kumar and Go, Ara and Choi, Han-Yong and Bang, Yunkyu},
  journal={New Journal of Physics},
  volume={22},
  number={6},
  pages={063048},
  year={2020},
  publisher={IOP Publishing},
  doi = {10.1088/1367-2630/ab84b7}
}

@article{Kuzian_PhysRevB.104.085154,
  title = {Ab initio based ligand field approach to determine electronic multiplet properties},
  author = {Kuzian, R. O. and Janson, O. and Savoyant, A. and van den Brink, Jeroen and Hayn, R.},
  journal = {Phys. Rev. B},
  volume = {104},
  issue = {8},
  pages = {085154},
  numpages = {17},
  year = {2021},
  month = {Aug},
  publisher = {American Physical Society},
  doi = {10.1103/PhysRevB.104.085154},
  url = {https://link.aps.org/doi/10.1103/PhysRevB.104.085154}
}

@article{matt2018direct,
  title={Direct observation of orbital hybridisation in a cuprate superconductor},
  author={Matt, Christian E and Sutter, D and Cook, AM and Sassa, Yasmine and M{\aa}nsson, Martin and Tjernberg, Oscar and Das, L and Horio, M and Destraz, D and Fatuzzo, CG and others},
  journal={Nature communications},
  volume={9},
  number={1},
  pages={972},
  year={2018},
  publisher={Nature Publishing Group UK London},
   doi = {10.1038/s41467-018-03266-0}
}

@article{Comin_PhysRevB.99.045105,
  title = {Resolving the nature of electronic excitations in resonant inelastic x-ray scattering},
  author = {Kang, M. and Pelliciari, J. and Krockenberger, Y. and Li, J. and McNally, D. E. and Paris, E. and Liang, R. and Hardy, W. N. and Bonn, D. A. and Yamamoto, H. and Schmitt, T. and Comin, R.},
  journal = {Phys. Rev. B},
  volume = {99},
  issue = {4},
  pages = {045105},
  numpages = {12},
  year = {2019},
  month = {Jan},
  publisher = {American Physical Society},
  doi = {10.1103/PhysRevB.99.045105},
  url = {https://link.aps.org/doi/10.1103/PhysRevB.99.045105}
}

@article{Barantani_PhysRevX.12.021068,
  title = {{Resonant Inelastic X-Ray Scattering Study of Electron-Exciton Coupling in High-${T}_{c}$ Cuprates}},
  author = {Barantani, F. and Tran, M. K. and Madan, I. and Kapon, I. and Bachar, N. and Asmara, T. C. and Paris, E. and Tseng, Y. and Zhang, W. and Hu, Y. and Giannini, E. and Gu, G. and Devereaux, T. P. and Berthod, C. and Carbone, F. and Schmitt, T. and van der Marel, D.},
  journal = {Phys. Rev. X},
  volume = {12},
  issue = {2},
  pages = {021068},
  numpages = {14},
  year = {2022},
  month = {Jun},
  publisher = {American Physical Society},
  doi = {10.1103/PhysRevX.12.021068},
  url = {https://link.aps.org/doi/10.1103/PhysRevX.12.021068}
}

@article{Sakakibara_2010,
  title = {{Two-Orbital Model Explains the Higher Transition Temperature of the Single-Layer Hg-Cuprate Superconductor Compared to That of the La-Cuprate Superconductor}},
  author = {Sakakibara, Hirofumi and Usui, Hidetomo and Kuroki, Kazuhiko and Arita, Ryotaro and Aoki, Hideo},
  journal = {Phys. Rev. Lett.},
  volume = {105},
  issue = {5},
  pages = {057003},
  numpages = {4},
  year = {2010},
  month = {Jul},
  publisher = {American Physical Society},
  doi = {10.1103/PhysRevLett.105.057003},
  url = {https://link.aps.org/doi/10.1103/PhysRevLett.105.057003}
}

@article{Sakakibara_2012,
  title = {{Origin of the material dependence of ${T}_{c}$ in the single-layered cuprates}},
  author = {Sakakibara, Hirofumi and Usui, Hidetomo and Kuroki, Kazuhiko and Arita, Ryotaro and Aoki, Hideo},
  journal = {Phys. Rev. B},
  volume = {85},
  issue = {6},
  pages = {064501},
  numpages = {12},
  year = {2012},
  month = {Feb},
  publisher = {American Physical Society},
  doi = {10.1103/PhysRevB.85.064501},
  url = {https://link.aps.org/doi/10.1103/PhysRevB.85.064501}
}

@article{Lee_PhysRevB.89.041104,
  title = {{Charge-orbital-lattice coupling effects in the $dd$ excitation profile of one-dimensional cuprates}},
  author = {Lee, J. J. and Moritz, B. and Lee, W. S. and Yi, M. and Jia, C. J. and Sorini, A. P. and Kudo, K. and Koike, Y. and Zhou, K. J. and Monney, C. and Strocov, V. and Patthey, L. and Schmitt, T. and Devereaux, T. P. and Shen, Z. X.},
  journal = {Phys. Rev. B},
  volume = {89},
  issue = {4},
  pages = {041104},
  numpages = {5},
  year = {2014},
  month = {Jan},
  publisher = {American Physical Society},
  doi = {10.1103/PhysRevB.89.041104},
  url = {https://link.aps.org/doi/10.1103/PhysRevB.89.041104}
}

@article{ivashko2019strain,
  title={{Strain-engineering mott-insulating La$_2$CuO$_4$}},
  author={Ivashko, Oleh and Horio, Masafumi and Wan, W and Christensen, NB and McNally, DE and Paris, E and Tseng, Y and Shaik, NE and R{\o}nnow, HM and Wei, HI and others},
  journal={Nature communications},
  volume={10},
  number={1},
  pages={786},
  year={2019},
  publisher={Nature Publishing Group UK London},
   doi = {10.1038/s41467-019-08664-6}
}

@article{Goodenough1955,
  author  = {J. B. Goodenough},
  title   = {{Theory of the Role of Covalence in the Perovskite-Type Manganites {[La, M(II)]MnO3}}},
  journal = {Physical Review},
  volume  = {100},
  pages   = {564--573},
  year    = {1955},
  doi     = {10.1103/PhysRev.100.564}
}

@article{Kanamori1959,
  author  = {J. Kanamori},
  title   = {Superexchange interaction and symmetry properties of electron orbitals},
  journal = {Journal of Physics and Chemistry of Solids},
  volume  = {10},
  pages   = {87--98},
  year    = {1959},
  doi     = {10.1016/0022-3697(59)90061-7}
}

@article{Anderson1959,
  author  = {P. W. Anderson},
  title   = {{New Approach to the Theory of Superexchange Interactions}},
  journal = {Physical Review},
  volume  = {115},
  pages   = {2--13},
  year    = {1959},
  doi     = {10.1103/PhysRev.115.2}
}

@article{Murakami_PhysRevLett.81.582,
  title = {{Resonant X-Ray Scattering from Orbital Ordering in ${\mathrm{LaMnO}}_{3}$}},
  author = {Murakami, Y. and Hill, J. P. and Gibbs, D. and Blume, M. and Koyama, I. and Tanaka, M. and Kawata, H. and Arima, T. and Tokura, Y. and Hirota, K. and Endoh, Y.},
  journal = {Phys. Rev. Lett.},
  volume = {81},
  issue = {3},
  pages = {582--585},
  numpages = {0},
  year = {1998},
  month = {Jul},
  publisher = {American Physical Society},
  doi = {10.1103/PhysRevLett.81.582},
  url = {https://link.aps.org/doi/10.1103/PhysRevLett.81.582}
}

@article{Pavarini_PhysRevLett.87.047003,
  title = {{Band-Structure Trend in Hole-Doped Cuprates and Correlation with ${\mathit{T}}_{\mathit{c}\mathrm{max}}$}},
  author = {Pavarini, E. and Dasgupta, I. and Saha-Dasgupta, T. and Jepsen, O. and Andersen, O. K.},
  journal = {Phys. Rev. Lett.},
  volume = {87},
  issue = {4},
  pages = {047003},
  numpages = {4},
  year = {2001},
  month = {Jul},
  publisher = {American Physical Society},
  doi = {10.1103/PhysRevLett.87.047003},
  url = {https://link.aps.org/doi/10.1103/PhysRevLett.87.047003}
}

@article{bogdanov2022enhancement,
  title={Enhancement of superexchange due to synergetic breathing and hopping in corner-sharing cuprates},
  author={Bogdanov, Nikolay A and Li Manni, Giovanni and Sharma, Sandeep and Gunnarsson, Olle and Alavi, Ali},
  journal={Nature Physics},
  volume={18},
  number={2},
  pages={190--195},
  year={2022},
  publisher={Nature Publishing Group UK London},
  doi = {10.1038/s41567-021-01439-1}
}

@article{little2007determination,
  title={{A determination of the pairing interaction in the high Tc cuprate superconductor Tl$_2$Ba$_2$CaCu$_2$O$_8$ (Tl2212)}},
  author={Little, WA and Holcomb, MJ and Ghiringhelli, G and Braicovich, Lucio and Dallera, Claudia and Piazzalunga, Andrea and Tagliaferri, Alberto and Brookes, NB},
  journal={Physica C: Superconductivity and its applications},
  volume={460},
  pages={40--43},
  year={2007},
  publisher={Elsevier},
  doi = {10.1016/j.physc.2007.03.031}
}

@article{cox1989virtual,
  title={{Virtual electric quadrupole fluctuations: a mechanism for high $T_c$}},
  author={Cox, DL and Jarrell, M and Jayaprakash, C and Krishna-murthy, Hulikal Ramaiyengar and Deisz, J},
  journal={Physical review letters},
  volume={62},
  number={18},
  pages={2188},
  year={1989},
  publisher={APS},
  doi = {10.1103/PhysRevLett.62.2188}
}

@article{hozoi2011ab,
  title={{Ab Initio determination of Cu 3 d orbital energies in layered copper oxides}},
  author={Hozoi, Liviu and Siurakshina, Liudmila and Fulde, Peter and van den Brink, Jeroen},
  journal={Scientific reports},
  volume={1},
  number={1},
  pages={65},
  year={2011},
  publisher={Nature Publishing Group UK London},
  doi = {10.1038/srep00065}
}

@article{Rossi_PhysRevB.104.L220505,
  title = {Orbital and spin character of doped carriers in infinite-layer nickelates},
  author = {Rossi, M. and Lu, H. and Nag, A. and Li, D. and Osada, M. and Lee, K. and Wang, B. Y. and Agrestini, S. and Garcia-Fernandez, M. and Kas, J. J. and Chuang, Y.-D. and Shen, Z. X. and Hwang, H. Y. and Moritz, B. and Zhou, Ke-Jin and Devereaux, T. P. and Lee, W. S.},
  journal = {Phys. Rev. B},
  volume = {104},
  issue = {22},
  pages = {L220505},
  numpages = {7},
  year = {2021},
  month = {Dec},
  publisher = {American Physical Society},
  doi = {10.1103/PhysRevB.104.L220505},
  url = {https://link.aps.org/doi/10.1103/PhysRevB.104.L220505}
}

@article{Timoshenko_PhysRevLett.120.225502,
  title = {{Neural Network Approach for Characterizing Structural Transformations by X-Ray Absorption Fine Structure Spectroscopy}},
  author = {Timoshenko, Janis and Anspoks, Andris and Cintins, Arturs and Kuzmin, Alexei and Purans, Juris and Frenkel, Anatoly I.},
  journal = {Phys. Rev. Lett.},
  volume = {120},
  issue = {22},
  pages = {225502},
  numpages = {6},
  year = {2018},
  month = {May},
  publisher = {American Physical Society},
  doi = {10.1103/PhysRevLett.120.225502},
  url = {https://link.aps.org/doi/10.1103/PhysRevLett.120.225502}
}

@article{Rankine2020,
author = {Rankine, C. D. and Madkhali, M. M. M. and Penfold, T. J.},
title = {A Deep Neural Network for the Rapid Prediction of X-ray Absorption Spectra},
journal = {The Journal of Physical Chemistry A},
volume = {124},
number = {21},
pages = {4263-4270},
year = {2020},
doi = {10.1021/acs.jpca.0c03723}
}

@article{corriero2023crystalmela,
  title={{CrystalMELA: a new crystallographic machine learning platform for crystal system determination}},
  author={Corriero, Nicola and Rizzi, Rosanna and Settembre, Gaetano and Del Buono, Nicoletta and Diacono, Domenico},
  journal={Applied Crystallography},
  volume={56},
  number={2},
  pages={409--419},
  year={2023},
  publisher={International Union of Crystallography},
  doi = {10.1107/S1600576723000596}
}

@article{jung2023automatic,
  title={Automatic materials characterization from infrared spectra using convolutional neural networks},
  author={Jung, Guwon and Jung, Son Gyo and Cole, Jacqueline M},
  journal={Chemical Science},
  volume={14},
  number={13},
  pages={3600--3609},
  year={2023},
  publisher={Royal Society of Chemistry},
  doi = {10.1039/D2SC05892H}
}

@article{schmid2023deconvolution,
  title={{Deconvolution of 1D NMR spectra: A deep learning-based approach}},
  author={Schmid, Nicolas and Bruderer, Simon and Paruzzo, F and Fischetti, G and Toscano, Giuseppe and Graf, Dominik and Fey, Michael and Henrici, Andreas and Ziebart, V and Heitmann, B and others},
  journal={Journal of Magnetic Resonance},
  volume={347},
  pages={107357},
  year={2023},
  publisher={Elsevier},
  doi = {10.1016/j.jmr.2022.107357}
}

@article{fan2019deep,
  title={Deep learning-based component identification for the Raman spectra of mixtures},
  author={Fan, Xiaqiong and Ming, Wen and Zeng, Huitao and Zhang, Zhimin and Lu, Hongmei},
  journal={Analyst},
  volume={144},
  number={5},
  pages={1789--1798},
  year={2019},
  publisher={Royal Society of Chemistry},
   doi = {10.1039/C8AN02212G}
}

@misc{tensorflow2015-whitepaper,
title={ {TensorFlow}: Large-Scale Machine Learning on Heterogeneous Systems},
url={https://www.tensorflow.org/},
note={Software available from tensorflow.org},
author={
    Mart\'{i}n~Abadi and
    Ashish~Agarwal and
    Paul~Barham and
    Eugene~Brevdo and
    Zhifeng~Chen and
    Craig~Citro and
    Greg~S.~Corrado and
    Andy~Davis and
    Jeffrey~Dean and
    Matthieu~Devin and
    Sanjay~Ghemawat and
    Ian~Goodfellow and
    Andrew~Harp and
    Geoffrey~Irving and
    Michael~Isard and
    Yangqing Jia and
    Rafal~Jozefowicz and
    Lukasz~Kaiser and
    Manjunath~Kudlur and
    Josh~Levenberg and
    Dandelion~Man\'{e} and
    Rajat~Monga and
    Sherry~Moore and
    Derek~Murray and
    Chris~Olah and
    Mike~Schuster and
    Jonathon~Shlens and
    Benoit~Steiner and
    Ilya~Sutskever and
    Kunal~Talwar and
    Paul~Tucker and
    Vincent~Vanhoucke and
    Vijay~Vasudevan and
    Fernanda~Vi\'{e}gas and
    Oriol~Vinyals and
    Pete~Warden and
    Martin~Wattenberg and
    Martin~Wicke and
    Yuan~Yu and
    Xiaoqiang~Zheng},
  year={2015},
}

@article{PhysRevLett.68.2543,
  title = {{Out-of-plane orbital characters of intrinsic and doped holes in ${\mathrm{La}}_{2\mathrm{\ensuremath{-}}\mathit{x}}$${\mathrm{Sr}}_{\mathit{x}}$${\mathrm{CuO}}_{4}$}},
  author = {Chen, C. T. and Tjeng, L. H. and Kwo, J. and Kao, H. L. and Rudolf, P. and Sette, F. and Fleming, R. M.},
  journal = {Phys. Rev. Lett.},
  volume = {68},
  issue = {16},
  pages = {2543--2546},
  numpages = {0},
  year = {1992},
  month = {Apr},
  publisher = {American Physical Society},
  doi = {10.1103/PhysRevLett.68.2543},
  url = {https://link.aps.org/doi/10.1103/PhysRevLett.68.2543}
}

@article{arpaia2023signature,
  title={{Signature of quantum criticality in cuprates by charge density fluctuations}},
  author={Arpaia, Riccardo and Martinelli, Leonardo and Sala, Marco Moretti and Caprara, Sergio and Nag, Abhishek and Brookes, Nicholas B and Camisa, Pietro and Li, Qizhi and Gao, Qiang and Zhou, Xingjiang and others},
  journal={Nature Communications},
  volume={14},
  number={1},
  pages={7198},
  year={2023},
  publisher={Nature Publishing Group UK London},
  doi = {10.1038/s41467-023-42961-5}
}

@article{PhysRevB.107.134513,
  title = {{Precise $dd$ excitations and commensurate intersite Coulomb interactions in the dissimilar cuprates $\mathrm{Y}{\mathrm{Ba}}_{2}{\mathrm{Cu}}_{3}{\mathrm{O}}_{7--y}$ and ${\mathrm{La}}_{2--x}{\mathrm{Sr}}_{x}\mathrm{Cu}{\mathrm{O}}_{4}$}},
  author = {Huang, Shih-Wen and Wray, L. Andrew and Shao, Yu-Cheng and Wu, Cheng-Yau and Wang, Shun-Hung and Lee, Jenn-Min and Chen, Y.-J. and Schoenlein, R. W. and Mou, C. Y. and Chuang, Yi-De and Lin, J.-Y.},
  journal = {Phys. Rev. B},
  volume = {107},
  issue = {13},
  pages = {134513},
  numpages = {8},
  year = {2023},
  month = {Apr},
  publisher = {American Physical Society},
  doi = {10.1103/PhysRevB.107.134513},
  url = {https://link.aps.org/doi/10.1103/PhysRevB.107.134513}
}

@article{lecun1998gradient,
  title={{Gradient-based learning applied to document recognition}},
  author={LeCun, Yann and Bottou, Leon and Bengio, Yoshua and Haffner, Patrick},
  journal={Proceedings of the IEEE},
  volume={86},
  number={11},
  pages={2278--2324},
  year={1998},
  doi={10.1109/5.726791}
}

@inproceedings{bromley1993signature,
  title     = {Signature Verification using a {Siamese} Time Delay Neural Network},
  author    = {Bromley, Jane and Guyon, Isabelle and LeCun, Yann and S{\"a}ckinger, Eduard and Shah, Roopak},
  booktitle = {Advances in Neural Information Processing Systems},
  volume    = {6},
  year      = {1993},
  url       = {https://papers.neurips.cc/paper_files/paper/1993/file/288cc0ff022877bd3df94bc9360b9c5d-Paper.pdf}
}

@inproceedings{szegedy2015going,
  title={{Going deeper with convolutions}},
  author={Szegedy, Christian and Liu, Wei and Jia, Yangqing and Sermanet, Pierre and Reed, Scott and Anguelov, Dragomir and Erhan, Dumitru and Vanhoucke, Vincent and Rabinovich, Andrew},
  booktitle={Proceedings of the IEEE Conference on Computer Vision and Pattern Recognition},
  pages={1--9},
  year={2015},
  url = {https://doi.org/10.48550/arXiv.1409.4842}
}

@inproceedings{yu2016multiscale,
  title={Multi-scale context aggregation by dilated convolutions},
  author={Yu, Fisher and Koltun, Vladlen},
  booktitle={International Conference on Learning Representations},
  year={2016},
  url = {https://doi.org/10.48550/arXiv.1511.07122}
}

@inproceedings{nair2010relu,
  title     = {{Rectified Linear Units Improve Restricted Boltzmann Machines}},
  author    = {Nair, Vinod and Hinton, Geoffrey E.},
  booktitle = {Proceedings of the 27th International Conference on Machine Learning},
  pages     = {807--814},
  year      = {2010},
  url       = {https://www.cs.toronto.edu/~hinton/absps/reluICML.pdf}
}

@inproceedings{glorot2011rectifier,
  title     = {Deep Sparse Rectifier Neural Networks},
  author    = {Glorot, Xavier and Bordes, Antoine and Bengio, Yoshua},
  booktitle = {Proceedings of the Fourteenth International Conference on Artificial Intelligence and Statistics},
  series    = {Proceedings of Machine Learning Research},
  volume    = {15},
  pages     = {315--323},
  year      = {2011},
  url       = {https://proceedings.mlr.press/v15/glorot11a.html}
}

@article{huber1964robust,
  title   = {{Robust Estimation of a Location Parameter}},
  author  = {Huber, Peter J.},
  journal = {The Annals of Mathematical Statistics},
  volume  = {35},
  number  = {1},
  pages   = {73--101},
  year    = {1964},
  doi     = {10.1214/aoms/1177703732},
  url     = {https://projecteuclid.org/journals/annals-of-mathematical-statistics/volume-35/issue-1/Robust-Estimation-of-a-Location-Parameter/10.1214/aoms/1177703732.full}
}

@inproceedings{kendall2018multitask,
  title     = {{Multi-Task Learning Using Uncertainty to Weigh Losses for Scene Geometry and Semantics}},
  author    = {Kendall, Alex and Gal, Yarin and Cipolla, Roberto},
  booktitle = {Proceedings of the IEEE Conference on Computer Vision and Pattern Recognition},
  pages     = {7482--7491},
  year      = {2018},
  url       = {https://openaccess.thecvf.com/content_cvpr_2018/papers/Kendall_Multi-Task_Learning_Using_CVPR_2018_paper.pdf}
}

@article{Azzalini1985,
  author  = {Azzalini, Adelchi},
  title   = {{A Class of Distributions Which Includes the Normal Ones}},
  journal = {Scandinavian Journal of Statistics},
  year    = {1985},
  volume  = {12},
  pages   = {171--178},
  url     = {https://www.jstor.org/stable/4615982}
}

@misc{HC5221,
  author = {HC5221},
  howpublished = {\doi {10.15151/ESRF-ES-1106943771}}
}

@article{Dockery2025,
  author  = {Dockery, Adam and Minamisono, Kei and Ortiz Cortes, Alejandro and Rickey, Brooke},
  title   = {{Spectral Line-Shape in Collinear Laser Spectroscopy after Atomic Charge Exchange}},
  journal = {Spectrochimica Acta Part B: Atomic Spectroscopy},
  year    = {2025},
  volume  = {233},
  pages   = {107280},
  doi     = {10.1016/j.sab.2025.107280}
}

@article{fischetti2025deep,
  title={{A deep learning framework for multiplet splitting classification in 1H NMR}},
  author={Fischetti, Giulia and Schmid, Nicolas and Bruderer, Simon and Heitmann, Bj{\"o}rn and Henrici, Andreas and Scarso, Alessandro and Caldarelli, Guido and Wilhelm, Dirk},
  journal={Journal of Magnetic Resonance},
  volume={373},
  pages={107851},
  year={2025},
  publisher={Elsevier},
  doi     = {10.1016/j.jmr.2025.107851}
}

@article{tnqm-ttj3,
  title = {{Hamiltonian parameter inference from resonant inelastic x-ray scattering with active learning}},
  author = {Lajer, Marton K. and Dai, Xin and Barros, Kipton and Carbone, Matthew R. and Johnston, S. and Dean, M. P. M.},
  journal = {Phys. Rev. B},
  volume = {112},
  issue = {15},
  pages = {155167},
  numpages = {16},
  year = {2025},
  month = {Oct},
  publisher = {American Physical Society},
  doi = {10.1103/tnqm-ttj3},
  url = {https://link.aps.org/doi/10.1103/tnqm-ttj3}
}

@misc{HC4412,
 author = {HC4412},
  howpublished = {\doi {10.15151/ESRF-ES-433978160}}
}

\clearpage
\onecolumngrid

\begin{center}
{\large\bfseries Supplementary Material for}\\[0.5em]
{\large\bfseries Machine-learning test of the single-ion model for $dd$ excitations in cuprates}
\end{center}

\setcounter{figure}{0}
\renewcommand{\thefigure}{S\arabic{figure}}

\setcounter{table}{0}
\renewcommand{\thetable}{S\arabic{table}}

\section{\label{sec:level1}Extended Data on LCO-AF and YBCO-AF}

\begin{figure}
    \includegraphics[width=\columnwidth]{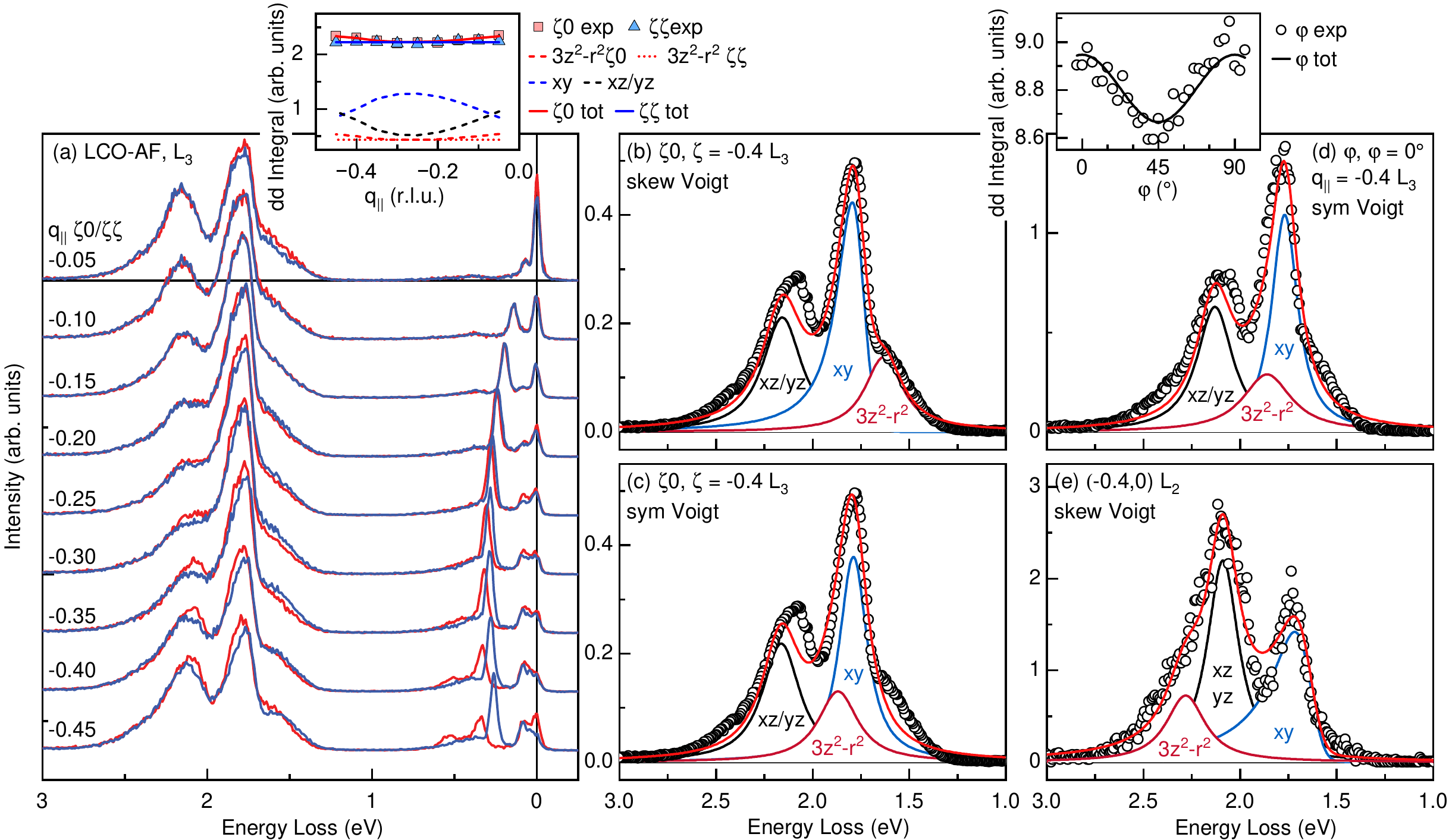}
    \caption{(a) RIXS spectra at the Cu L$_3$ edge for the LCO-AF
samples measured along the ($\zeta$,0) (red lines), ($\zeta$,$\zeta$) (blue lines) directions. All data are taken at $L = 1.3$\,r.l.u., where $L$ is the component of $\mathbf{q}$ along the $c$-direction. (b, c) Decomposition and fit of an Cu L$_3$ LCO-AF RIXS spectrum measured along ($\zeta$,0), using a skew-Voigt profile in (b) and a symmetric Voigt profile in (c). (d) Same as in (c), but for a spectrum from the $\varphi$ scan. (e) Decomposition and fit of a LCO-AF RIXS spectrum measured at ($-0.4,0$) and at the Cu L$_2$ edge. The insets in (a) and (d) show $q_{\parallel}$ and $\varphi$-dependence of the $dd$ excitation cross sections at the Cu L$_3$ edge calculated within the single-ion model (solid and dashed lines) with markers indicating the energy-integrated intensity of the raw experimental spectra between 1 and 3\,eV. }
    \label{fig:fig_s1}
\end{figure}

For LCO-AF, we repeated the global fitting procedure by modeling the $xy$ peak as a symmetric Voigt profile. The results, shown in Fig.\,\ref{fig:fig_s1}, demonstrate that the position of the $3z^2-r^2$ peak is highly sensitive to the choice of lineshape for the $xy$ peak. As an additional example, Fig.\,\ref{fig:fig_s1}\,(e) shows a Cu L$_2$ edge spectrum fitted with the single-ion model, where the $3z^2-r^2$ excitation is found at $2.28$\,eV. However, fitting a single spectrum alone is not reliable, as more robust information is obtained from the $\mathbf{q}$-dependence.\par

\begin{figure}
    \includegraphics[width=\columnwidth]{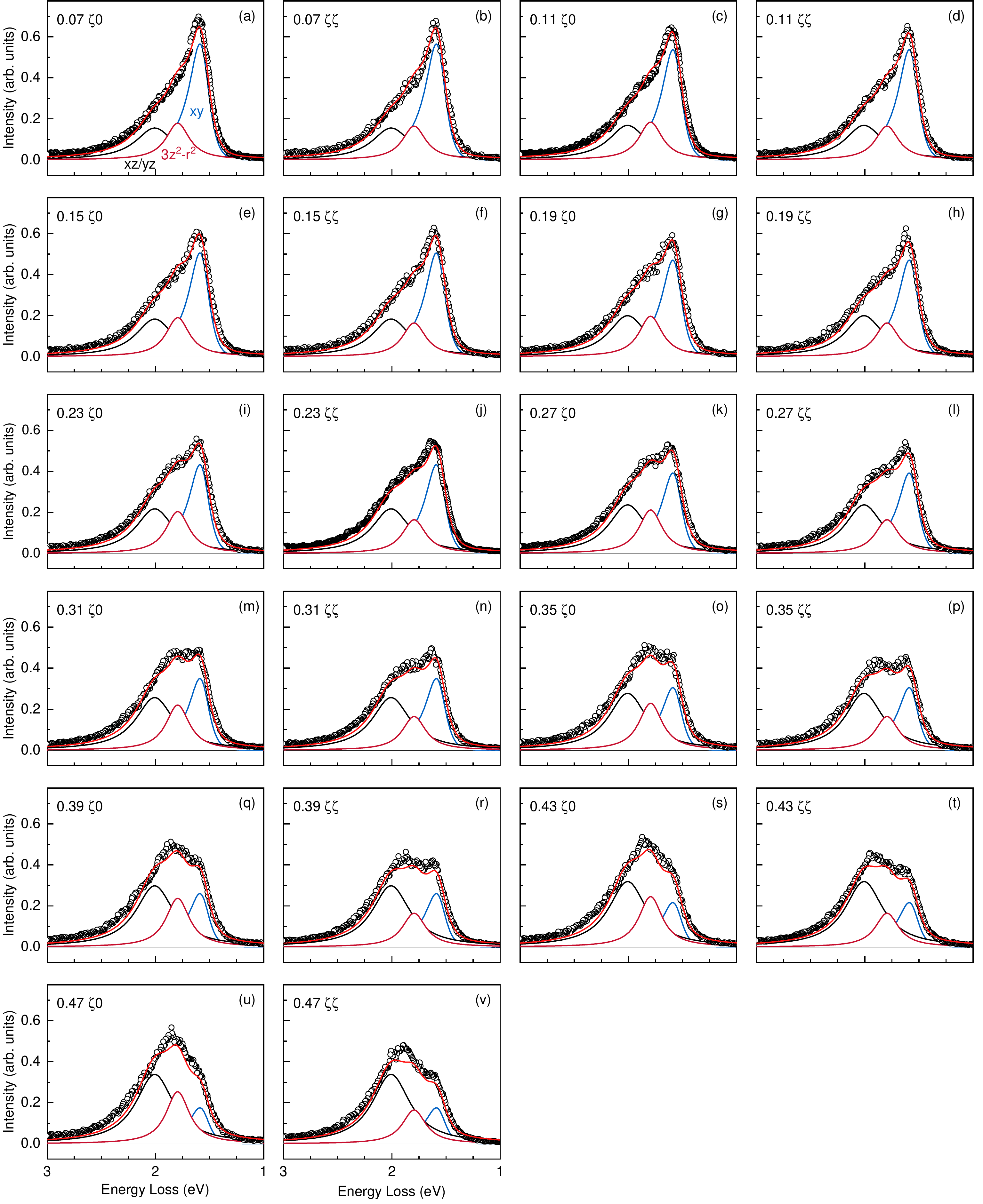}
    \caption{Global fit of all RIXS spectra of YBCO-AF along the ($\zeta$,0) and ($\zeta$,$\zeta$) directions. All data are taken at $2\theta=149.5^{\circ}$. }
    \label{fig:fig_s2}
\end{figure}

\begin{figure}
    \includegraphics[width=\columnwidth]{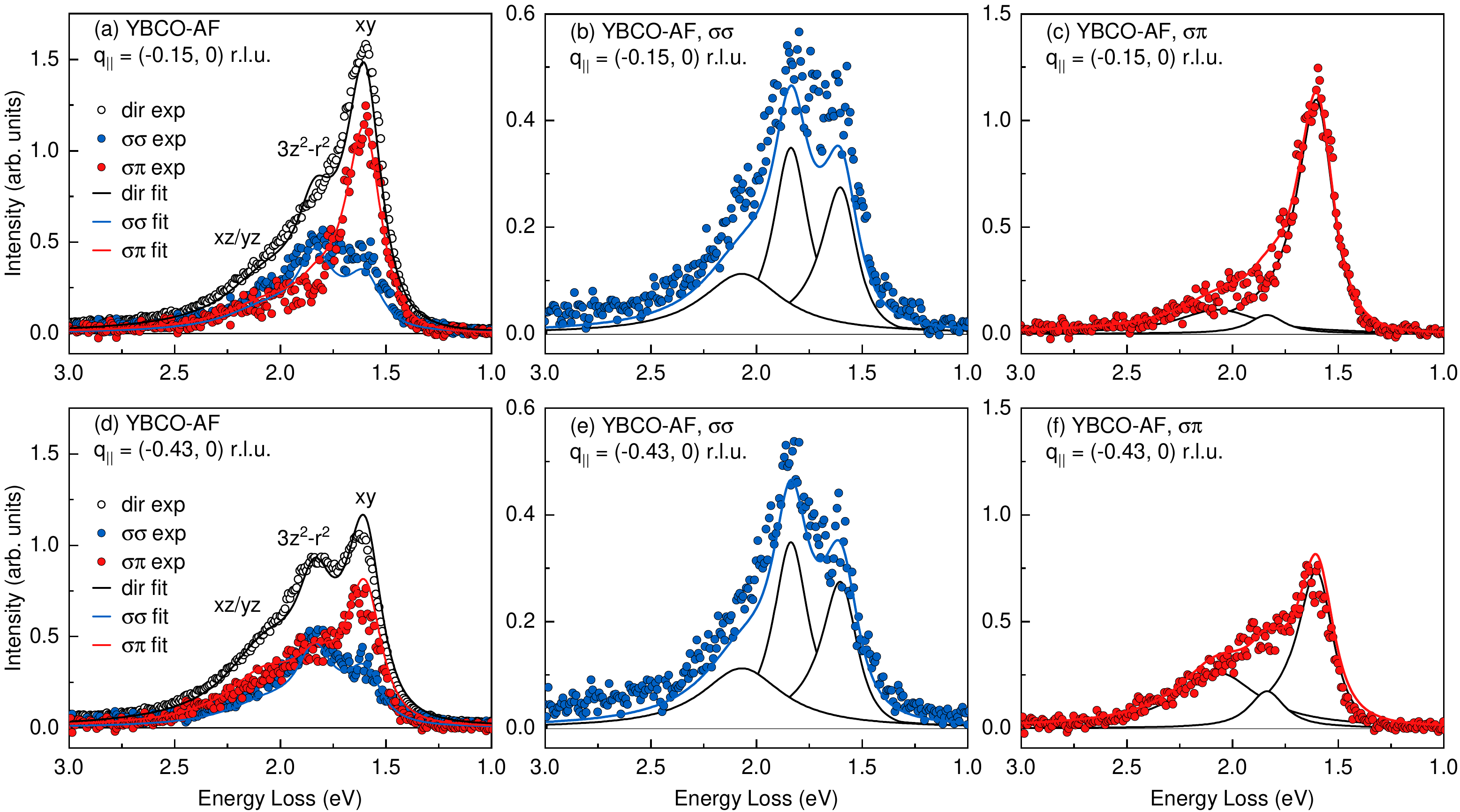}
    \caption{Decomposition and fit of the polarization-resolved YBCO-AF spectra at (a–c) $\mathbf{q}_{\parallel}=(-0.15,0)$ and (d–f) $\mathbf{q}_{\parallel}=(-0.43,0)$\,r.l.u.. Blue spectra correspond to the non-crossed polarization channels ($\sigma\sigma$, $\pi\pi$), whereas red spectra correspond to the crossed polarization channels ($\sigma\pi$, $\pi\sigma$). All data are taken at $2\theta=149.5^{\circ}$.}
    \label{fig:fig_s3}
\end{figure}

Figure\,\ref{fig:fig_s2} displays the global fit of the YBCO-AF spectra measured along the $(\zeta, 0)$ and $(\zeta, \zeta)$ directions, confirming the same peak assignment as in the $\varphi$ scan. This assignment is further supported by the global fitting of the polarization-resolved measurements at $\mathbf{q}_{\parallel}=(-0.15,0)$ and $\mathbf{q}_{\parallel}=(-0.43,0)$\,r.l.u., shown in Fig.\,\ref{fig:fig_s3}. Table\,\ref{tab:summary_all} summarizes all parameters obtained from the global fitting for YBCO-AF and LCO-AF.\par

The skew-Voigt profile used for the fitting is defined as
\begin{equation}
I_{\mathrm{sv}}(\omega)=
A\,
\frac{
2\,V(\omega;\omega_0,\gamma,\sigma)\,
\Phi\!\left[\alpha\,\frac{\omega-\omega_0}{\sigma}\right]
}{
\displaystyle
\int_{-\infty}^{+\infty}
2\,V(\omega';\omega_0,\gamma,\sigma)\,
\Phi\!\left[\alpha\,\frac{\omega'-\omega_0}{\sigma}\right]\,
d\omega'
}
\end{equation}

\begin{equation}
\Phi(x)=\frac{1}{2}\left[1+\operatorname{erf}\!\left(\frac{x}{\sqrt{2}}\right)\right]
\end{equation}

\begin{equation}
V(\omega;\omega_0,\gamma,\sigma)
=
\frac{\Re[w(z)]}{\sigma\sqrt{2\pi}},
\qquad
z=\frac{(\omega-\omega_0)+i\gamma}{\sqrt{2}\sigma}
\end{equation}

Here, $A$ is the integrated intensity, $\omega_0$ is the peak position, $\gamma$ is the Lorentzian half-width at half-maximum, $\sigma$ is the Gaussian standard deviation, $\alpha$ is the asymmetry parameter, $\Re[w(z)]$ is the real part of the Faddeeva function $w(z)$, and $\Phi(x)$ is the cumulative distribution function of the standard normal distribution.

\begin{table*}[t]
\caption{
Summary of peak positions ($E$), Lorentzian HWHM ($L$), experimental resolution (Res), and total FWHM ($W$) for the YBCO-AF and LCO-AF samples obtained from global fitting. All values except $\alpha$ (dimensionless) are in eV. The error bars are one to three orders of magnitude smaller than the fitting results and are therefore omitted.}

\label{tab:summary_all}
\centering
\small
\setlength{\tabcolsep}{5pt}

\renewcommand{\arraystretch}{1.15}
\resizebox{\textwidth}{!}{
\begin{tabular}{ll c c c c c c c c c c c}
\toprule
Sample & Scan
& \multicolumn{1}{c}{$E_{3z^2-r^2}$}
& \multicolumn{1}{c}{$E_{xy}$}
& \multicolumn{1}{c}{$E_{xz/yz}$}
& \multicolumn{1}{c}{$L_{3z^2-r^2}$}
& \multicolumn{1}{c}{$L_{xy}$}
& \multicolumn{1}{c}{$L_{xz/yz}$}
& \multicolumn{1}{c}{Res}
& \multicolumn{1}{c}{$\alpha$}
& \multicolumn{1}{c}{$W_{3z^2-r^2}$}
& \multicolumn{1}{c}{$W_{xy}$}
& \multicolumn{1}{c}{$W_{xz/yz}$} \\
\midrule
YBCO-AF & $\varphi$
& 1.83
& 1.56
& 2.00
& 0.10
& 0.11
& 0.19
& 0.04
& 0.20
& 0.21
& 0.20
& 0.38\\

YBCO-AF & $(\zeta,0)$, $(\zeta,\zeta)$
& 1.79
& 1.56
& 2.00
& 0.13
& 0.11
& 0.21
& 0.045
& 0.12
& 0.26
& 0.23
& 0.42 \\

YBCO-AF & $(-0.15,0)$, $(-0.43,0)$
& 1.84
& 1.59
& 2.08
& 0.10
& 0.10
& 0.23
& 0.04
& 0.05
& 0.20
& 0.21
& 0.48\\
\midrule
LCO-AF & $\varphi$
& 1.63
& 1.72
& 2.13
& 0.10
& 0.12
& 0.11
& 0.04
& 0.41
& 0.21
& 0.17 
& 0.23  \\

LCO-AF & $(\zeta,0)$, $(\zeta,\zeta)$
& 1.63
& 1.75
& 2.15
& 0.11
& 0.09
& 0.12
& 0.045
& 0.28
& 0.23
& 0.17
& 0.25\\

LCO-AF & L$_2$ 1 spectrum
& 2.28
& 1.64
& 2.08
& 0.12
& 0.20
& 0.09
& 0.04
& 0.29
& 0.24
& 0.26
& 0.20\\
\bottomrule
\end{tabular}
}
\end{table*}

\section{\label{sec:level2}Neural network performances on synthetic spectra}

Before applying the neural network to experimental spectra, we evaluated its performance on synthetic data generated within the single-ion model. The complete synthetic dataset contained 300,000 synthetic samples and was randomly split into 80\% for training and 20\% for validation/testing, corresponding to 240,000 and 60,000 samples, respectively. The 60,000 held-out samples were not used to update the network weights, but only to monitor the validation loss during training. The performance evaluated on this held-out set quantifies how accurately the neural network predicts the known peak parameters for spectra generated by the same model adopted for training.\par

\begin{figure}
    \includegraphics[width=\columnwidth]{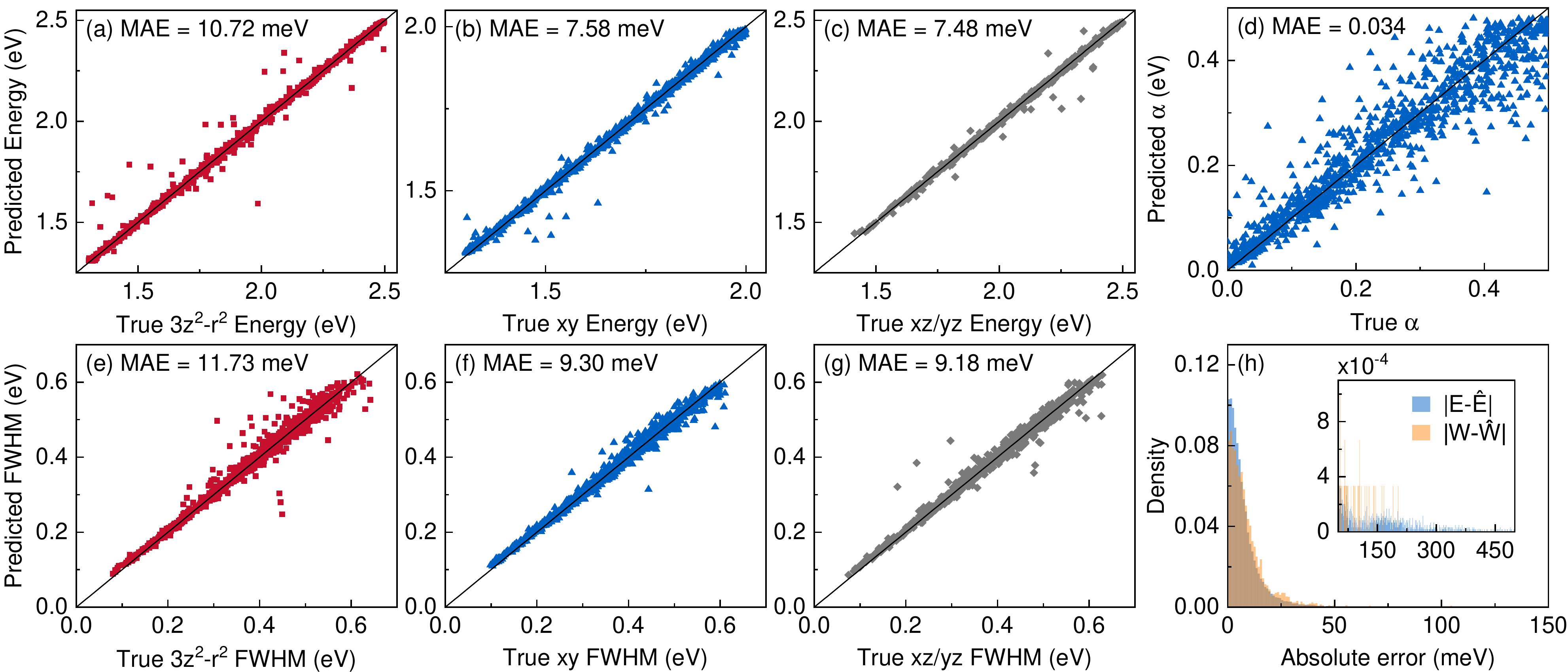}
    \caption{Neural-network performance on synthetic data generated within the single-ion model. For clarity, the scatter plots include only 1000 randomly selected examples, while the mean absolute errors (MAE) are calculated over all 60,000 held-out samples. (a–c) Predicted versus true excitation energies for the $3z^2-r^2$, $xy$, and $xz/yz$ orbital channels. (d) Predicted versus true $xy$ asymmetry parameter $\alpha$. (e–g) Predicted versus true total FWHM values for the same three channels.  In (a–g), the black solid line corresponds to $\hat{y}=y$, i.e. exact agreement between prediction and target value. (h) Absolute-error distributions for the excitation energies and FWHM values. The inset highlights the low-density tails of the distributions.}
    \label{fig:fig_s_new}
\end{figure}

The results are summarized in Fig.\,\ref{fig:fig_s_new}. Panels (a-c) compare the true and predicted energies for the three orbital excitations. The points closely follow the diagonal line, showing that the network accurately predicts the peak positions over the full energy range of the synthetic dataset. The mean absolute errors (MAE), defined as $|E-\hat{E}|$, where $E$ is the true value and $\hat{E}$ the predicted value, are 10.72\,meV, 7.58\,meV, and 7.48\,meV for the $3z^2-r^2$, $xy$, and $xz/yz$ orbitals, respectively. Panels (e-g) show the corresponding comparison for the total FWHM values. The linewidths are also well reproduced, although with a slightly broader scatter than the peak energies. The mean absolute errors are 11.73\,meV, 9.30\,meV, and 9.18\,meV for the $3z^2-r^2$, $xy$, and $xz/yz$ channels, respectively. Panel (d) shows the predicted versus true values of the $xy$ asymmetry parameter, with a mean absolute error of 0.034. The larger scatter shows that $\alpha$ is harder for the network to determine than the excitation energies and linewidths, likely because broad or overlapping peaks make the line shape less distinct.\par

Finally, panel (h) shows the absolute-error distributions for the excitation energies and FWHM values. Both distributions are strongly peaked near zero and decay rapidly, confirming that the network accurately predicts the target parameters for most spectra in the held-out set. The low-density tails correspond to difficult cases, such as strongly overlapping peaks, large linewidths, or geometries in which the orbital intensity contrast is weak.\par

Overall, this synthetic validation shows that the neural network can reliably extract the peak parameters from spectra generated within the single-ion framework. Therefore, discrepancies or unstable assignments observed when the same network is applied to experimental spectra should not be attributed only to a failure of the regression algorithm, but may instead indicate that the experimental spectra contain contributions not included in the single-ion model.

\section{\label{sec:level2}Global fitting performance on synthetic spectra}

We analyzed the synthetic spectra using the same global fitting strategy adopted for the experimental data. The procedure was applied to 1000 synthetic samples. For each sample, the four spectra were fitted simultaneously, sharing three excitation energies, three Lorentzian linewidths, and an asymmetry parameter $\alpha$. The Gaussian contribution was fixed to the resolution value used in the simulation, which was common to the four spectra of each sample but differed from sample to sample. The relative $dd$ intensities were calculated within the single-ion model using the stored scattering geometry and incident polarization, and a single global scale factor was optimized for each sample.\par

\begin{figure}
    \includegraphics[width=\columnwidth]{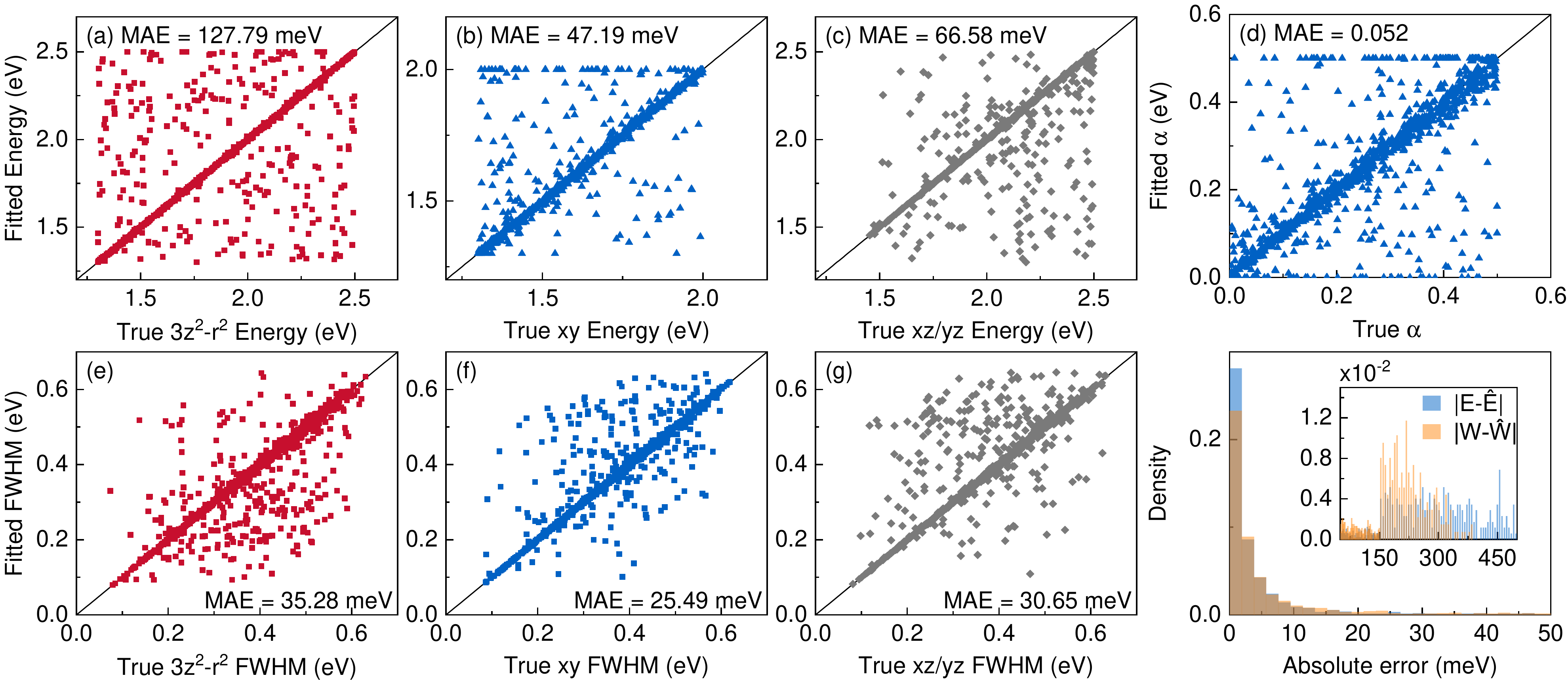}
    \caption{Global fitting performance on 1000 synthetic samples used for the training of the neural network. (a–c) Fitted versus true excitation energies for the $3z^2-r^2$, $xy$, and $xz/yz$ orbital channels. (d) Fitted versus true $xy$ asymmetry parameter $\alpha$. (e–g) Fitted versus true total FWHM values for the same three channels.  In (a–g), the black solid line corresponds to $\hat{y}=y$, i.e. exact agreement between fitted and target value. (h) Absolute-error distributions for the excitation energies and FWHM values. The inset highlights the low-density tails of the distributions.}
    \label{fig:fig_s_new2}
\end{figure}

For each sample, the minimization was repeated from three different initial conditions. One start used excitation energies of 1.3\,eV, Lorentzian widths of 0.1\,eV, and $\alpha=0$, while the other two starts were initialized randomly within the allowed parameter ranges. The solution with the lowest cost function was retained. The parameters were constrained by broad bounds: 1.3--2.5\,eV for the excitation energies of $3z^2-r^2$ and $xz/yz$, 1.3--2\,eV for the excitation energies of $xy$, 0.03--0.25\,eV for the Lorentzian HWHM, and 0--0.5 for $\alpha$. The results of the global fitting on the synthetic data are shown in Fig.\,\ref{fig:fig_s_new2}. The global fit converged for all tested samples, but the extracted parameters show a substantially larger scatter than the neural-network predictions in Fig.\,\ref{fig:fig_s_new}. The mean absolute errors on the excitation energies are 127.79, 47.19, and 66.58\,meV for the $3z^2-r^2$, $xy$, and $xz/yz$ orbitals, respectively. The corresponding errors on the total FWHM values are 35.28, 25.49, and 30.65\,meV, while the mean absolute error on the $\alpha$ parameter is 0.052. This lower accuracy is likely related to the limited number of spectra included in each fit. Global fitting is more stable when many spectra are fitted simultaneously, whereas in this case only four spectra are available for each sample. As a result, the fit becomes more sensitive to the choice of initial parameters, local minima, and ambiguities due to overlapping peaks, compared with the neural network approach.\par

\section{\label{sec:level2}Neural network performances on real spectra}

\begin{figure}
    \includegraphics[width=\columnwidth]{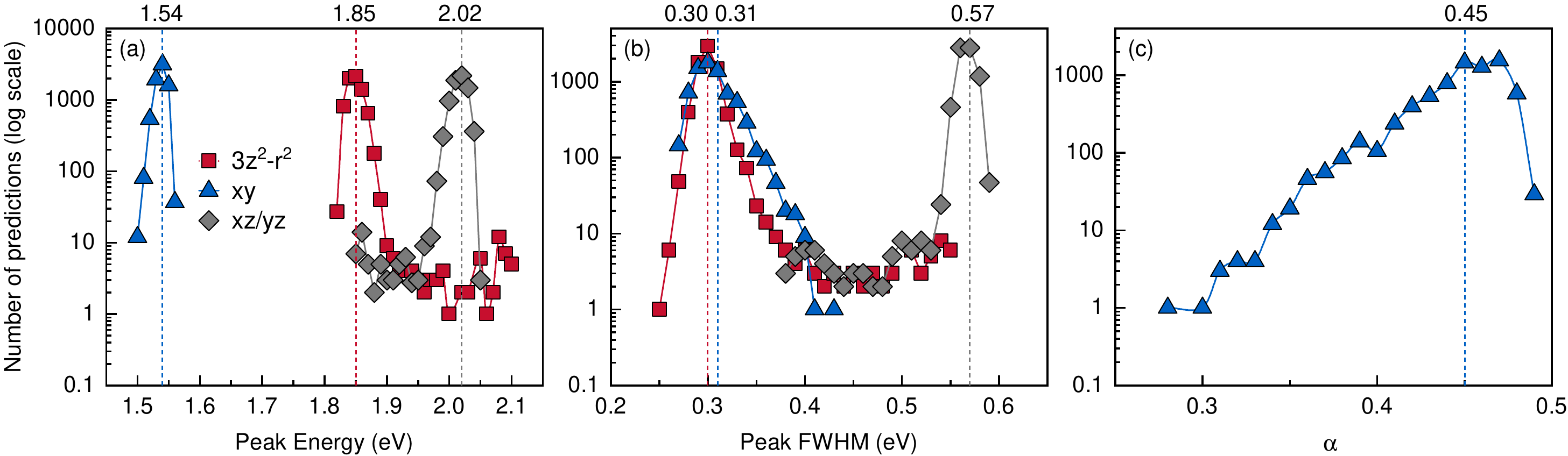}
    \caption{Summary plots of the three-variant NN predicted (a) peak positions and (b) widths and (c) $\alpha$ of the YBCO-AF spectra taken along the ($\zeta$, 0) and ($\zeta$, $\zeta$) directions. The dashed lines indicate the mean value of each parameter. The data correspond to model no.\,8 in Table\,\ref{tab:nn_architecture_comparison}.}
    \label{fig:fig_s4}
\end{figure}

After validating the neural network on synthetic spectra, we tested the robustness of the final three-variant model on experimental spectra.  We applied the network to the YBCO-AF data measured along the  ($\zeta$,0) and ($\zeta$,$\zeta$) directions, for a total of 18 scans. Since the neural network takes 4 scans as input at a time, the 18 scans were grouped into all possible subsets of 4. This gives 7315 distinct combinations. To test whether the network is sensitive to the ordering of the input spectra, we also considered all 24 possible permutations of the 4 scans within each subset. The neural network was then applied to every combination and every permutation, and the corresponding predictions were recorded.\par
Since the network architecture is not strictly permutation-invariant, different orderings of the same 4 inputs can produce slight numerical differences in the output. Nevertheless, the largest variation observed was smaller than $4 \times 10^{-4}$, indicating that this dependence on input ordering is negligible in practice.\par

Figure\,\ref{fig:fig_s4} shows the predictions obtained from all 7315 combinations on a logarithmic scale, with dashed lines indicating the mean values. The $xy$ excitation is always found between 1.5 and 1.6\,eV,  whereas the $3z^2-r^2$ and $xz/yz$ peaks are more broadly distributed around 1.84 and 2.0\,eV, respectively. This shows that, in a few cases, the ordering of the $3z^2-r^2$ and $xz/yz$ peaks is inverted. To evaluate the performance of the neural network, we counted the number of predictions lying outside the intervals $E_{3z^2-r^2}=1.85\pm0.05$\,eV, $E_{xy}=1.55\pm0.05$\,eV, and $E_{xz/yz}=2.00\pm0.05$\,eV. Across 10 neural networks trained on the same dataset, the number of out-of-range combinations varies between 76 and 247, which is less than 4\% of the 7315 total combinations. For the $\varphi$ scan, where only seven spectra are available, the neural network takes as input four spectra at different geometries and correctly predicts the corresponding values in all cases.\par

\section{\label{sec:level2}0 and one-variant Network Architectures}

\begin{figure}
    \includegraphics[width=\columnwidth]{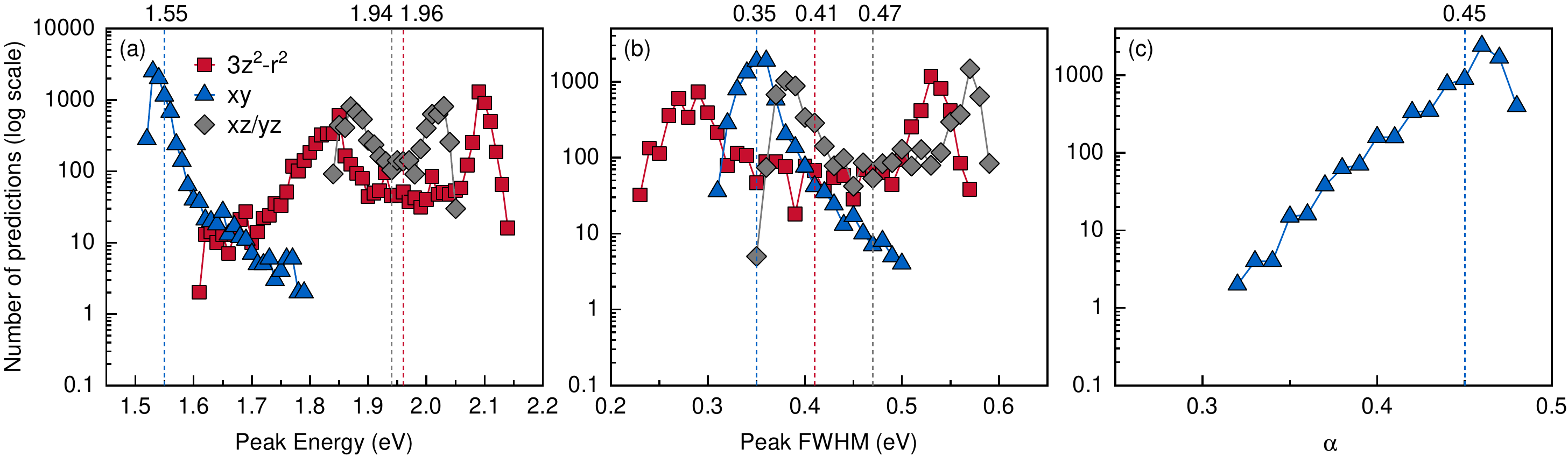}
    \caption{Summary plots of the 0-variant NN predicted (a) peak positions and (b) widths and (c) $\alpha$ of the YBCO-AF spectra taken along the ($\zeta$, 0) and ($\zeta$, $\zeta$) directions. The dashed lines indicate the mean value of each parameter. The data correspond to model no.\,1 in Table\,\ref{tab:nn_architecture_comparison}.}
    \label{fig:fig_s5}
\end{figure}

\begin{figure}
    \includegraphics[width=\columnwidth]{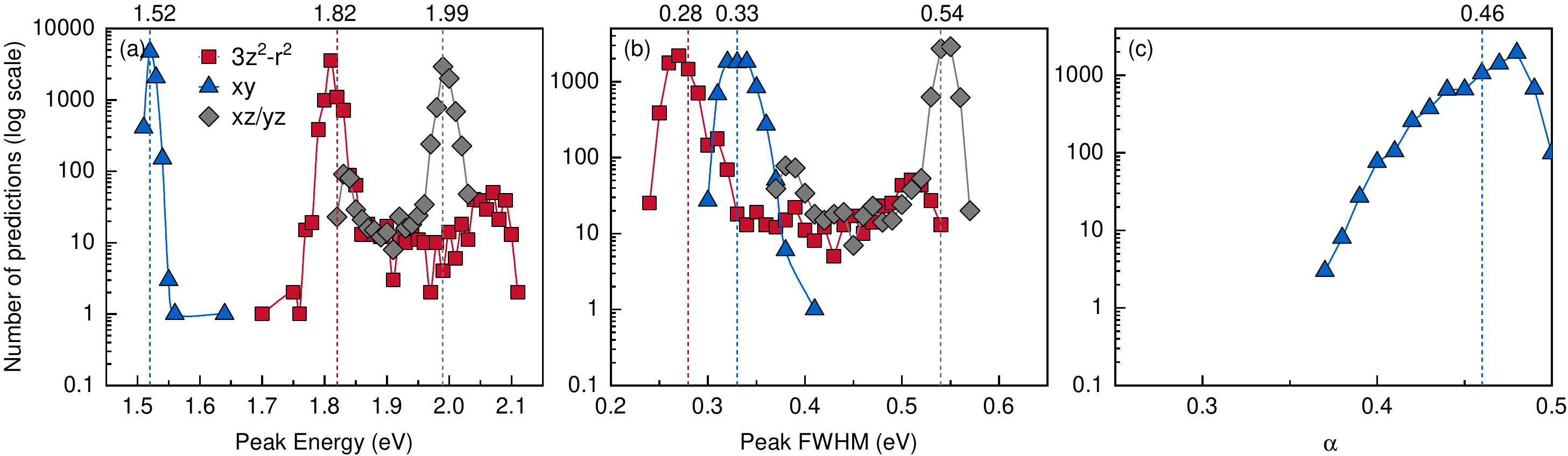}
    \caption{Summary plots of the one-variant NN predicted (a) peak positions and (b) widths and (c) $\alpha$ of the YBCO-AF spectra taken along the ($\zeta$, 0) and ($\zeta$, $\zeta$) directions. The dashed lines indicate the mean value of each parameter. The data correspond to model no.\,3 in Table\,\ref{tab:nn_architecture_comparison}.}
    \label{fig:fig_s6}
\end{figure}

Before introducing the three-variant architecture, we first considered a simpler network without preprocessing variants, 0-variant model. For each sample, the four input spectra were jointly normalized so that their total integrated intensity was unity and then ordered by increasing total $dd$ spectral weight. Residual degeneracies were removed by successive ordering according to the increasing areas of the $3z^2-r^2$, $xy$, and $xz/yz$ excitations and, if necessary, by the original geometry index. The network input included the four ordered spectra and the corresponding normalized $dd$ areas for the four geometries. This version therefore relied on a single canonical ordering, without difference spectra or complementary reordered views.\par

Next, we considered an intermediate network based on a single preprocessing variant, one-variant model. In this case, for each sample the $3z^2-r^2$ contribution was rescaled to have the same integrated intensity across the four geometries. The spectra were then ordered by increasing $xz/yz$ area, with residual degeneracies resolved by the increasing $xy$ area and, if necessary, by the original geometry index. Afterward, the four spectra were jointly normalized so that their total integrated intensity was unity. The network input included the ordered spectra, the corresponding difference spectra with respect to the first ordered geometry and the rescaled and normalized $dd$ areas. In both simplified architectures, the network followed the same multiscale convolutional strategy as the final model and predicted the three excitation energies, the three total FWHM values, and the $xy$ asymmetry parameter $\alpha$.\par

Figures\,\ref{fig:fig_s5} and \ref{fig:fig_s6} show summaries of the predictions obtained with the 0-variant and one-variant neural networks, respectively. In both cases, the predicted distributions are significantly broader than for the three-variant model, and in 0-variant approach the $xy$ and $3z^2-r^2$ peaks can be exchanged. Consistently, the number of predictions falling outside the defined above energy ranges increases to about 6000 and 2000 cases for the 0-variant and the one-variant neural networks, respectively. These results indicate that both simplified architectures are substantially less reliable than the final three-variant model, see Table \,\ref{tab:nn_architecture_comparison} for all details.\par

\begin{table}[t]
\caption{
Comparison of the 0-, one-, and three-variant neural-network architectures applied to the YBCO-AF spectra measured along the $(\zeta,0)$ and $(\zeta,\zeta)$ directions. 
For each architecture, the same experimental test was repeated using independently trained neural networks. In the main text and in the following, the results are shown for the three-variant model no.\,8.}
\label{tab:nn_architecture_comparison}
\centering
\small
\resizebox{\textwidth}{!}{
\renewcommand{\arraystretch}{1.2}
\begin{threeparttable}
\begin{tabular}{l c c c}
\toprule
Architecture 
& Trained NNs
& Out-of-range cases
& Out-of-range cases \% \\
\midrule
0-variant NN
& 3
& 4907, 6133, 6355
& 67.08, 83.84, 86.88 \\

One-variant NN
& 3
& 2480, 1434, 1168
& 33.90, 19.60, 15.97 \\

Three-variant NN
& 10
& 137, 116, 134, 95, 218, 83, 151, 76, 199, 247
& 1.04--3.38 \\
\bottomrule
\end{tabular}
\end{threeparttable}
}
\end{table}

\section{\label{sec:level3}NN results for LCO-AF ($\zeta$,0) and ($\zeta$,$\zeta$) scans}

\begin{figure}
    \includegraphics[width=\columnwidth]{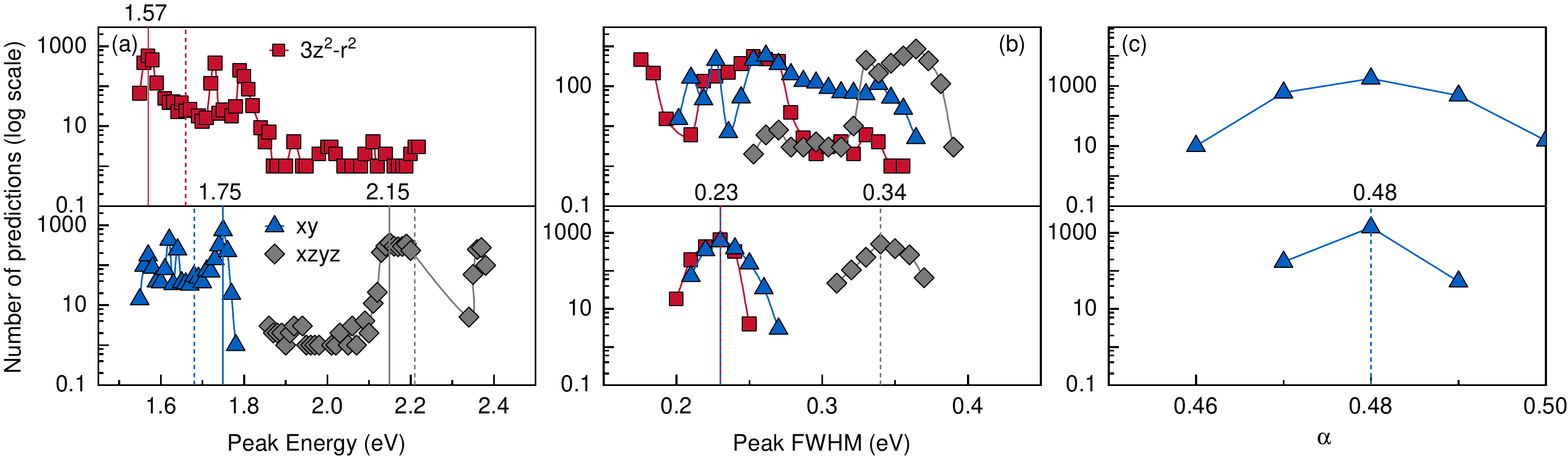}
    \caption{Summary plots of the three-variant NN predicted values (a) peak positions and (b, top) widths and (c, top) $\alpha$ of the LCO-AF spectra taken along the ($\zeta$, 0) and ($\zeta$, $\zeta$) directions.  In (a), the solid lines and numbers above indicate the peak value of each $dd$ excitation energy, while the dashed lines indicate the corresponding mean values. Panels (b, bottom) and (c, bottom) report the predicted FWHM and $\alpha$ values for cases in which the predicted peak energies fall within $\pm 0.05$\,eV of the corresponding peak values. The dashed lines indicate the mean value of each parameter.}
    \label{fig:fig_s7}
\end{figure}

Figure\,\ref{fig:fig_s7}\,(a) shows the $dd$ energy predictions for LCO-AF obtained from all 7315 combinations on a logarithmic scale. The solid lines indicate the most probable values, while the dashed lines indicate the mean values. Since the peak energies exhibit a broad distribution, we used the most probable value of each energy in our analysis. To obtain the corresponding FWHM from the full set (see Fig.\,\ref{fig:fig_s7}\,(b), top), we selected only predictions for which $E_{3z^2-r^2}$, $E_{xy}$, and $E_{xz/yz}$ lie within $\pm 0.05$\,eV of their respective most probable energies. From this subset, we extracted the corresponding FWHM values and computed their mean values, see Fig.\,\ref{fig:fig_s7}\,(b), bottom. The same procedure was applied to the asymmetry parameter $\alpha$, as shown in Fig.\,\ref{fig:fig_s7}\,(c). The most probable energy values, as well as the mean FWHM and $\alpha$ values obtained in this way, were used for Fig.\,4 and Table\,I in the main text.\par

\section{Single-ion calculations}\label{SIM}

In this section, we summarize the analytical expressions used to evaluate the polarization-resolved $dd$ excitation cross sections within the single-ion model for $x^2-y^2$ ground state. We first introduce the scattering geometry, the polarization vectors, and the spin-dependent quantities. Next, we list the orbital scattering matrices for spin-orbit coupling (SOC) $=0$ and the explicit expressions for the intensities in all polarization channels.

\subsection{Scattering geometry}
In RIXS $\theta$ is the angle between the incident x-ray beam and the sample surface, while $2\theta$ is the total scattering angle between the incident and outgoing beams ($2\theta\neq2 \times \theta$). With this convention, the incidence and exit angles measured with respect to the sample surface are
\begin{equation}
\theta_{\mathrm{in}} = \theta - \frac{\pi}{2}.
\qquad
\theta_{\mathrm{out}} = \theta - 2\theta + \frac{\pi}{2},
\end{equation}
Therefore,
\begin{equation}
\cos\theta_{\mathrm{in}} = \sin \theta,
\qquad
\sin\theta_{\mathrm{in}} = -\cos \theta,
\end{equation}
\begin{equation}
\cos\theta_{\mathrm{out}} = \sin(2\theta-\theta),
\qquad
\sin\theta_{\mathrm{out}} = \cos(2\theta-\theta).
\end{equation}

\subsection{Polarization vectors}

The same azimuthal angle $\varphi$ is
\begin{equation}
\varphi_{\mathrm{in}}=\varphi_{\mathrm{out}}=\varphi.
\end{equation}
The polarization vectors are then
\begin{equation}
\hat{\sigma}=
\begin{pmatrix}
-\sin\varphi\\
\cos\varphi\\
0
\end{pmatrix},
\end{equation}
\begin{equation}
\hat{\pi}_{\mathrm{in}}=
\begin{pmatrix}
-\cos\theta_{\mathrm{in}}\cos\varphi\\
-\cos\theta_{\mathrm{in}}\sin\varphi\\
\sin\theta_{\mathrm{in}}
\end{pmatrix},
\qquad
\hat{\pi}_{\mathrm{out}}=
\begin{pmatrix}
-\cos\theta_{\mathrm{out}}\cos\varphi\\
-\cos\theta_{\mathrm{out}}\sin\varphi\\
\sin\theta_{\mathrm{out}}
\end{pmatrix}.
\end{equation}

\subsection{Spin-dependent quantities}

The spin direction is specified by the polar angle $\theta_{\mathrm{spin}}$ and the azimuthal angle $\varphi_{\mathrm{spin}}$. The corresponding Cartesian components are
\begin{equation}
n_x = \sin\theta_{\mathrm{spin}}\cos\varphi_{\mathrm{spin}},
\qquad
n_y = \sin\theta_{\mathrm{spin}}\sin\varphi_{\mathrm{spin}},
\qquad
n_z = \cos\theta_{\mathrm{spin}}.
\end{equation}
We also introduce
\begin{equation}
\chi = e^{-i\varphi_{\mathrm{spin}}},
\qquad
u = \cos\!\left(\frac{\theta_{\mathrm{spin}}}{2}\right),
\qquad
v = \sin\!\left(\frac{\theta_{\mathrm{spin}}}{2}\right),
\end{equation}
so that
\begin{equation}
\gamma = i\,\chi\,\sin\theta_{\mathrm{spin}},
\qquad
\alpha = u^2 + \chi^2 v^2,
\qquad
\beta = u^2 - \chi^2 v^2.
\end{equation}
Equivalently,
\begin{equation}
\gamma = i\,e^{-i\varphi_{\mathrm{spin}}}\sin\theta_{\mathrm{spin}},
\end{equation}
\begin{equation}
\alpha =
\cos^2\!\left(\frac{\theta_{\mathrm{spin}}}{2}\right)
+
e^{-2i\varphi_{\mathrm{spin}}}
\sin^2\!\left(\frac{\theta_{\mathrm{spin}}}{2}\right),
\end{equation}
\begin{equation}
\beta =
\cos^2\!\left(\frac{\theta_{\mathrm{spin}}}{2}\right)
-
e^{-2i\varphi_{\mathrm{spin}}}
\sin^2\!\left(\frac{\theta_{\mathrm{spin}}}{2}\right).
\end{equation}

\subsection{General expression}

For each orbital block $M_j$, the scattering amplitude in the polarization channel
$p_{\mathrm{in}}\to p_{\mathrm{out}}$ is
\begin{equation}
A_j^{(p_{\mathrm{in}}\to p_{\mathrm{out}})}
=
\mathbf e_{\mathrm{in}}^{T} M_j \mathbf e_{\mathrm{out}},
\end{equation}
where $\mathbf e_{\mathrm{in}}$ is either $\hat{\sigma}$ or $\hat{\pi}_{\mathrm{in}}$, while $\mathbf e_{\mathrm{out}}$ is either $\hat{\sigma}$ or $\hat{\pi}_{\mathrm{out}}$. The corresponding cross section is
\begin{equation}
I_j^{(p_{\mathrm{in}}\to p_{\mathrm{out}})}
=
\left|
A_j^{(p_{\mathrm{in}}\to p_{\mathrm{out}})}
\right|^2.
\end{equation}

\subsection{$dd$ scattering matrices for SOC$=0$}

The $3\times 3$ orbital blocks entering the calculation are

\begin{equation}
M_{x^2-y^2,\mathrm{nsf}}=
\begin{pmatrix}
2 & i n_z & 0\\
-i n_z & 2 & 0\\
0&0&0
\end{pmatrix},
\qquad
M_{x^2-y^2,\mathrm{sf}}=
\gamma
\begin{pmatrix}
0 & -1 & 0\\
1 & 0 & 0\\
0&0&0
\end{pmatrix},
\end{equation}

\begin{equation}
M_{z^2,\mathrm{nsf}}=
\frac{1}{\sqrt{3}}
\begin{pmatrix}
-2 & i n_z & 2 i n_y\\
i n_z & 2 & 2 i n_x\\
0&0&0
\end{pmatrix},
\qquad
M_{z^2,\mathrm{sf}}=
\frac{1}{\sqrt{3}}
\begin{pmatrix}
0 & -\gamma & 2\alpha\\
-\gamma & 0 & 2 i \beta\\
0&0&0
\end{pmatrix},
\end{equation}

\begin{equation}
M_{xy,\mathrm{nsf}}=
\begin{pmatrix}
- i n_z & 2 & 0\\
-2 & - i n_z & 0\\
0&0&0
\end{pmatrix},
\qquad
M_{xy,\mathrm{sf}}=
\gamma
\begin{pmatrix}
1 & 0 & 0\\
0 & 1 & 0\\
0&0&0
\end{pmatrix},
\end{equation}

\begin{equation}
M_{yz,\mathrm{nsf}}=
\begin{pmatrix}
- i n_y & 0 & -2\\
- i n_x & 0 & i n_z\\
0&0&0
\end{pmatrix},
\qquad
M_{yz,\mathrm{sf}}=
\begin{pmatrix}
-\alpha & 0 & 0\\
- i\beta & 0 & -\gamma\\
0&0&0
\end{pmatrix},
\end{equation}

\begin{equation}
M_{xz,\mathrm{nsf}}=
\begin{pmatrix}
0 & - i n_y & i n_z\\
0 & - i n_x & 2\\
0&0&0
\end{pmatrix},
\qquad
M_{xz,\mathrm{sf}}=
\begin{pmatrix}
0 & -\alpha & -\gamma\\
0 & - i\beta & 0\\
0&0&0
\end{pmatrix}.
\end{equation}

\subsection{Explicit cross sections}

In the following, for each orbital channel we report the explicit expressions of the intensities in the four polarization geometries $\sigma\to\sigma$, $\sigma\to\pi$, $\pi\to\sigma$, and $\pi\to\pi$.

$x^2-y^2_{\mathrm{nsf}}$
\begin{align}
I_{x^2-y^2,\mathrm{nsf}}^{\sigma\to\sigma}
&=4,\\
I_{x^2-y^2,\mathrm{nsf}}^{\sigma\to\pi}
&=
\cos^2\theta_{\mathrm{out}}\,\cos^2\theta_{\mathrm{spin}},\\
I_{x^2-y^2,\mathrm{nsf}}^{\pi\to\sigma}
&=
\cos^2\theta_{\mathrm{in}}\,\cos^2\theta_{\mathrm{spin}},\\
I_{x^2-y^2,\mathrm{nsf}}^{\pi\to\pi}
&=
4\cos^2\theta_{\mathrm{in}}\cos^2\theta_{\mathrm{out}}.
\end{align}

$x^2-y^2_{\mathrm{sf}}$
\begin{align}
I_{x^2-y^2,\mathrm{sf}}^{\sigma\to\sigma}
&=0,\\
I_{x^2-y^2,\mathrm{sf}}^{\sigma\to\pi}
&=
\sin^2\theta_{\mathrm{spin}}\,\cos^2\theta_{\mathrm{out}},\\
I_{x^2-y^2,\mathrm{sf}}^{\pi\to\sigma}
&=
\sin^2\theta_{\mathrm{spin}}\,\cos^2\theta_{\mathrm{in}},\\
I_{x^2-y^2,\mathrm{sf}}^{\pi\to\pi}
&=0.
\end{align}

$3z^2-r^2_{\mathrm{nsf}}$
\begin{align}
I_{z^2,\mathrm{nsf}}^{\sigma\to\sigma}
&=
\frac{1}{3}
\left(
4\cos^2 2\varphi + n_z^2 \sin^2 2\varphi
\right),\\
I_{z^2,\mathrm{nsf}}^{\sigma\to\pi}
&=
\frac{1}{3}
\left[
4\cos^2\theta_{\mathrm{out}}\sin^2 2\varphi
+
\left(
2\sin\theta_{\mathrm{out}}
\left(
n_x\cos\varphi-n_y\sin\varphi
\right)
-
\cos\theta_{\mathrm{out}}\,n_z\cos 2\varphi
\right)^2
\right],\\
I_{z^2,\mathrm{nsf}}^{\pi\to\sigma}
&=
\frac{\cos^2\theta_{\mathrm{in}}}{3}
\left(
4\sin^2 2\varphi + n_z^2 \cos^2 2\varphi
\right),\\
I_{z^2,\mathrm{nsf}}^{\pi\to\pi}
&=
\frac{\cos^2\theta_{\mathrm{in}}}{3}
\left[
4\cos^2\theta_{\mathrm{out}}\cos^2 2\varphi
+
\left(
\cos\theta_{\mathrm{out}}\,n_z\sin 2\varphi
-
2\sin\theta_{\mathrm{out}}
\left(
n_x\sin\varphi+n_y\cos\varphi
\right)
\right)^2
\right].
\end{align}

$3z^2-r^2_{\mathrm{sf}}$

\begin{align}
I_{z^2,\mathrm{sf}}^{\sigma\to\sigma}
&=
\frac{1}{3}\sin^2\theta_{\mathrm{spin}}\,\sin^2 2\varphi,\\
I_{z^2,\mathrm{sf}}^{\sigma\to\pi}
&=
\frac{1}{3}
\left|
\cos\theta_{\mathrm{out}}\,\gamma \cos 2\varphi
+
2\sin\theta_{\mathrm{out}}
\left(
-\alpha\sin\varphi+i\beta\cos\varphi
\right)
\right|^2,\\
I_{z^2,\mathrm{sf}}^{\pi\to\sigma}
&=
\frac{\cos^2\theta_{\mathrm{in}}}{3}
\sin^2\theta_{\mathrm{spin}}\,\cos^2 2\varphi,\\
I_{z^2,\mathrm{sf}}^{\pi\to\pi}
&=
\frac{\cos^2\theta_{\mathrm{in}}}{3}
\left|
\cos\theta_{\mathrm{out}}\,\gamma \sin 2\varphi
+
2\sin\theta_{\mathrm{out}}
\left(
\alpha\cos\varphi+i\beta\sin\varphi
\right)
\right|^2.
\end{align}

$xy_{\mathrm{sf}}$
\begin{align}
I_{xy,\mathrm{sf}}^{\sigma\to\sigma}
&=
\sin^2\theta_{\mathrm{spin}},\\
I_{xy,\mathrm{sf}}^{\sigma\to\pi}
&=0,\\
I_{xy,\mathrm{sf}}^{\pi\to\sigma}
&=0,\\
I_{xy,\mathrm{sf}}^{\pi\to\pi}
&=
\sin^2\theta_{\mathrm{spin}}\,
\cos^2\theta_{\mathrm{in}}\,
\cos^2\theta_{\mathrm{out}}.
\end{align}

$xy_{\mathrm{nsf}}$
\begin{align}
I_{xy,\mathrm{nsf}}^{\sigma\to\sigma}
&=
\cos^2\theta_{\mathrm{spin}},\\
I_{xy,\mathrm{nsf}}^{\sigma\to\pi}
&=
4\cos^2\theta_{\mathrm{out}},\\
I_{xy,\mathrm{nsf}}^{\pi\to\sigma}
&=
4\cos^2\theta_{\mathrm{in}},\\
I_{xy,\mathrm{nsf}}^{\pi\to\pi}
&=
\cos^2\theta_{\mathrm{spin}}\,
\cos^2\theta_{\mathrm{in}}\,
\cos^2\theta_{\mathrm{out}}.
\end{align}

$yz_{\mathrm{nsf}}$
\begin{align}
I_{yz,\mathrm{nsf}}^{\sigma\to\sigma}
&=
\sin^2\varphi
\left(
n_x\cos\varphi-n_y\sin\varphi
\right)^2,\\
I_{yz,\mathrm{nsf}}^{\sigma\to\pi}
&=
4\sin^2\theta_{\mathrm{out}}\sin^2\varphi
+
\cos^2\varphi
\left(
\cos\theta_{\mathrm{out}}
\left(
n_x\cos\varphi-n_y\sin\varphi
\right)
+
\sin\theta_{\mathrm{out}}\,n_z
\right)^2,\\
I_{yz,\mathrm{nsf}}^{\pi\to\sigma}
&=
\cos^2\theta_{\mathrm{in}}\,
\sin^2\varphi
\left(
n_x\sin\varphi+n_y\cos\varphi
\right)^2,\\
I_{yz,\mathrm{nsf}}^{\pi\to\pi}
&=
\cos^2\theta_{\mathrm{in}}
\left[
4\sin^2\theta_{\mathrm{out}}\cos^2\varphi
+
\left(
\cos\theta_{\mathrm{out}}\cos\varphi
\left(
n_x\sin\varphi+n_y\cos\varphi
\right)
+
\sin\theta_{\mathrm{out}}\,n_z\sin\varphi
\right)^2
\right].
\end{align}

$yz_{\mathrm{sf}}$
\begin{align}
I_{yz,\mathrm{sf}}^{\sigma\to\sigma}
&=
\sin^2\varphi
\left|
-\alpha\sin\varphi+i\beta\cos\varphi
\right|^2,\\
I_{yz,\mathrm{sf}}^{\sigma\to\pi}
&=
\cos^2\varphi
\left|
\cos\theta_{\mathrm{out}}
\left(
-\alpha\sin\varphi+i\beta\cos\varphi
\right)
-
\sin\theta_{\mathrm{out}}\,\gamma
\right|^2,\\
I_{yz,\mathrm{sf}}^{\pi\to\sigma}
&=
\cos^2\theta_{\mathrm{in}}\,
\sin^2\varphi
\left|
\alpha\cos\varphi+i\beta\sin\varphi
\right|^2,\\
I_{yz,\mathrm{sf}}^{\pi\to\pi}
&=
\cos^2\theta_{\mathrm{in}}
\left|
\cos\theta_{\mathrm{out}}\cos\varphi
\left(
\alpha\cos\varphi+i\beta\sin\varphi
\right)
-
\sin\theta_{\mathrm{out}}\sin\varphi\,\gamma
\right|^2.
\end{align}

$xz_{\mathrm{nsf}}$
\begin{align}
I_{xz,\mathrm{nsf}}^{\sigma\to\sigma}
&=
\cos^2\varphi
\left(
n_x\cos\varphi-n_y\sin\varphi
\right)^2,\\
I_{xz,\mathrm{nsf}}^{\sigma\to\pi}
&=
4\sin^2\theta_{\mathrm{out}}\cos^2\varphi
+
\sin^2\varphi
\left(
\cos\theta_{\mathrm{out}}
\left(
n_x\cos\varphi-n_y\sin\varphi
\right)
-
\sin\theta_{\mathrm{out}}\,n_z
\right)^2,\\
I_{xz,\mathrm{nsf}}^{\pi\to\sigma}
&=
\cos^2\theta_{\mathrm{in}}\,
\cos^2\varphi
\left(
n_x\sin\varphi+n_y\cos\varphi
\right)^2,\\
I_{xz,\mathrm{nsf}}^{\pi\to\pi}
&=
\cos^2\theta_{\mathrm{in}}
\left[
4\sin^2\theta_{\mathrm{out}}\sin^2\varphi
+
\left(
\cos\theta_{\mathrm{out}}\sin\varphi
\left(
n_x\sin\varphi+n_y\cos\varphi
\right)
+
\sin\theta_{\mathrm{out}}\,n_z\cos\varphi
\right)^2
\right].
\end{align}

$xz_{\mathrm{sf}}$
\begin{align}
I_{xz,\mathrm{sf}}^{\sigma\to\sigma}
&=
\cos^2\varphi
\left|
-\alpha\sin\varphi+i\beta\cos\varphi
\right|^2,\\
I_{xz,\mathrm{sf}}^{\sigma\to\pi}
&=
\sin^2\varphi
\left|
\cos\theta_{\mathrm{out}}
\left(
-\alpha\sin\varphi+i\beta\cos\varphi
\right)
+
\sin\theta_{\mathrm{out}}\,\gamma
\right|^2,\\
I_{xz,\mathrm{sf}}^{\pi\to\sigma}
&=
\cos^2\theta_{\mathrm{in}}\,
\cos^2\varphi
\left|
\alpha\cos\varphi+i\beta\sin\varphi
\right|^2,\\
I_{xz,\mathrm{sf}}^{\pi\to\pi}
&=
\cos^2\theta_{\mathrm{in}}
\left|
\cos\theta_{\mathrm{out}}\sin\varphi
\left(
\alpha\cos\varphi+i\beta\sin\varphi
\right)
-
\sin\theta_{\mathrm{out}}\cos\varphi\,\gamma
\right|^2.
\end{align}

\subsection{Sums of spin-flip and non-spin-flip contributions}

For completeness, we also report the sum of the spin-flip and non-spin-flip contributions. 

\begin{equation}
I_j^{p_{\mathrm{in}}\to p_{\mathrm{out}}}
=
I_{j,\mathrm{nsf}}^{p_{\mathrm{in}}\to p_{\mathrm{out}}}
+
I_{j,\mathrm{sf}}^{p_{\mathrm{in}}\to p_{\mathrm{out}}},
\end{equation}
for all orbital channels and polarization geometries. Although each term individually depends on the spin orientation, this dependence cancels out in their sum. Consequently, the total cross section for each $dd$ excitation is independent of the spin direction.

$x^2-y^2$ 

\begin{align}
I_{x^2-y^2}^{\sigma\to\sigma}
&=
I_{x^2-y^2,\mathrm{nsf}}^{\sigma\to\sigma}
+
I_{x^2-y^2,\mathrm{sf}}^{\sigma\to\sigma}
=4,\\
I_{x^2-y^2}^{\sigma\to\pi}
&=
I_{x^2-y^2,\mathrm{nsf}}^{\sigma\to\pi}
+
I_{x^2-y^2,\mathrm{sf}}^{\sigma\to\pi}
=\cos^2\theta_{\mathrm{out}},\\
I_{x^2-y^2}^{\pi\to\sigma}
&=
I_{x^2-y^2,\mathrm{nsf}}^{\pi\to\sigma}
+
I_{x^2-y^2,\mathrm{sf}}^{\pi\to\sigma}
=\cos^2\theta_{\mathrm{in}},\\
I_{x^2-y^2}^{\pi\to\pi}
&=
I_{x^2-y^2,\mathrm{nsf}}^{\pi\to\pi}
+
I_{x^2-y^2,\mathrm{sf}}^{\pi\to\pi}
=
4\cos^2\theta_{\mathrm{in}}\cos^2\theta_{\mathrm{out}}.
\end{align}

$3z^2-r^2$ 

\begin{align}
I_{z^2}^{\sigma\to\sigma}
&=
I_{z^2,\mathrm{nsf}}^{\sigma\to\sigma}
+
I_{z^2,\mathrm{sf}}^{\sigma\to\sigma}
=
\frac{1}{3}
\left(
4\cos^2 2\varphi+\sin^2 2\varphi
\right),\\
I_{z^2}^{\sigma\to\pi}
&=
I_{z^2,\mathrm{nsf}}^{\sigma\to\pi}
+
I_{z^2,\mathrm{sf}}^{\sigma\to\pi}
=
\frac{1}{3}
\left(
4-\cos^2\theta_{\mathrm{out}}\cos^2 2\varphi
\right),\\
I_{z^2}^{\pi\to\sigma}
&=
I_{z^2,\mathrm{nsf}}^{\pi\to\sigma}
+
I_{z^2,\mathrm{sf}}^{\pi\to\sigma}
=
\frac{\cos^2\theta_{\mathrm{in}}}{3}
\left(
4\sin^2 2\varphi+\cos^2 2\varphi
\right),\\
I_{z^2}^{\pi\to\pi}
&=
I_{z^2,\mathrm{nsf}}^{\pi\to\pi}
+
I_{z^2,\mathrm{sf}}^{\pi\to\pi}
=
\frac{\cos^2\theta_{\mathrm{in}}}{3}
\left(
4-3\cos^2\theta_{\mathrm{out}}\sin^2 2\varphi
\right).
\end{align}

$xy$

\begin{align}
I_{xy}^{\sigma\to\sigma}
&=
I_{xy,\mathrm{nsf}}^{\sigma\to\sigma}
+
I_{xy,\mathrm{sf}}^{\sigma\to\sigma}
=1,\\
I_{xy}^{\sigma\to\pi}
&=
I_{xy,\mathrm{nsf}}^{\sigma\to\pi}
+
I_{xy,\mathrm{sf}}^{\sigma\to\pi}
=
4\cos^2\theta_{\mathrm{out}},\\
I_{xy}^{\pi\to\sigma}
&=
I_{xy,\mathrm{nsf}}^{\pi\to\sigma}
+
I_{xy,\mathrm{sf}}^{\pi\to\sigma}
=
4\cos^2\theta_{\mathrm{in}},\\
I_{xy}^{\pi\to\pi}
&=
I_{xy,\mathrm{nsf}}^{\pi\to\pi}
+
I_{xy,\mathrm{sf}}^{\pi\to\pi}
=
\cos^2\theta_{\mathrm{in}}\cos^2\theta_{\mathrm{out}}.
\end{align}

$yz$

\begin{align}
I_{yz}^{\sigma\to\sigma}
&=
I_{yz,\mathrm{nsf}}^{\sigma\to\sigma}
+
I_{yz,\mathrm{sf}}^{\sigma\to\sigma}
=
\sin^2\varphi,\\
I_{yz}^{\sigma\to\pi}
&=
I_{yz,\mathrm{nsf}}^{\sigma\to\pi}
+
I_{yz,\mathrm{sf}}^{\sigma\to\pi}
=
\cos^2\varphi
+
4\sin^2\theta_{\mathrm{out}}\sin^2\varphi,\\
I_{yz}^{\pi\to\sigma}
&=
I_{yz,\mathrm{nsf}}^{\pi\to\sigma}
+
I_{yz,\mathrm{sf}}^{\pi\to\sigma}
=
\cos^2\theta_{\mathrm{in}}\sin^2\varphi,\\
I_{yz}^{\pi\to\pi}
&=
I_{yz,\mathrm{nsf}}^{\pi\to\pi}
+
I_{yz,\mathrm{sf}}^{\pi\to\pi}
=
\cos^2\theta_{\mathrm{in}}
\left[
\cos^2\varphi
+
\left(1+2\cos^2\varphi\right)\sin^2\theta_{\mathrm{out}}
\right].
\end{align}

$xz$

\begin{align}
I_{xz}^{\sigma\to\sigma}
&=
I_{xz,\mathrm{nsf}}^{\sigma\to\sigma}
+
I_{xz,\mathrm{sf}}^{\sigma\to\sigma}
=
\cos^2\varphi,\\
I_{xz}^{\sigma\to\pi}
&=
I_{xz,\mathrm{nsf}}^{\sigma\to\pi}
+
I_{xz,\mathrm{sf}}^{\sigma\to\pi}
=
\sin^2\varphi
+
4\sin^2\theta_{\mathrm{out}}\cos^2\varphi,\\
I_{xz}^{\pi\to\sigma}
&=
I_{xz,\mathrm{nsf}}^{\pi\to\sigma}
+
I_{xz,\mathrm{sf}}^{\pi\to\sigma}
=
\cos^2\theta_{\mathrm{in}}\cos^2\varphi,\\
I_{xz}^{\pi\to\pi}
&=
I_{xz,\mathrm{nsf}}^{\pi\to\pi}
+
I_{xz,\mathrm{sf}}^{\pi\to\pi}
=
\cos^2\theta_{\mathrm{in}}
\left[
\sin^2\varphi
+
\left(1+2\sin^2\varphi\right)\sin^2\theta_{\mathrm{out}}
\right].
\end{align}

\end{document}